# Doubling the near-infrared photocurrent in a solar cell via omni-resonant coherent perfect absorption

Massimo L. Villinger[1], Abbas Shiri[1,2], Soroush Shabahang[1,3], Ali K. Jahromi[1], Magued B. Nasr[4], Christopher H. Villinger[1], and Ayman F. Abouraddy[1,5,*]

[1]CREOL, The College of Optics & Photonics, University of Central Florida, Orlando, FL 32816, USA

[2]Department of Electrical and Computer Engineering, University of Central Florida, Orlando, FL 32816, USA

[3]The Wellman Center for Photomedicine at Massachusetts General Hospital, Harvard Medical School, Cambridge, MA 02139, USA4

[4]Department of Physics, Boston University, Boston, MA 02215, USA

[5]Department of Materials Science and Engineering, University of Central Florida, Orlando, FL 32816, USA

*corresponding author: raddy@creol.ucf.edu

## Abstract

Minimizing the material usage in thin-film solar cells can reduce manufacturing costs and enable mechanically flexible implementations, but concomitantly diminishes optical absorption. Coherent optical effects can help alleviate this inevitable drawback at discrete frequencies. For example, coherent perfect absorption guarantees that light is fully absorbed in a thin layer regardless of material or thickness – but only on resonance. Here we show that 'omni-resonance' delivers such coherent enhancement over a broad bandwidth by structuring the optical field to nullify the angular dispersion intrinsic to resonant structures. After embedding an amorphous-silicon thin-film photovoltaic cell in a planar cavity, pre-conditioning the incident light using an alignment-free optical arrangement severs the link between the resonant bandwidth and the cavity-photon lifetime, thereby rendering the cavity omni-resonant. Coherently enhanced near-infrared absorption doubles the photocurrent over the targeted spectral range 660–740 nm where every wavelength resonates. These results may pave the way to transparent solar cells that optimally harvest near-infrared light.

# Introduction

The promise held by solar energy as a renewable resource has fueled investigations of all the aspects that may improve the performance of photovoltaic cells [1]. A variety of approaches have been pursued, including surface patterning [2] or modifying the local density of states [3] to enhance light trapping beyond the ray-optic limit [4], multi-junction solar cells [5], band-splitting [6], and exploiting plasmonic structures [7]. Thin-film solar-cell technologies lower the cost and weight by utilizing less material, and offer additional benefits of mechanical flexibility coupled with the potential for low-cost manufacturing on large foil substrates via roll- to-roll processing [8,9]. The inevitable diminishing of optical absorption associated with reduced-thickness solar cells can be overcome by exploiting coherent resonant optical effects that boost the absorption. For example, coherent perfect absorption (CPA) [10,11] ensures unconditional *full* absorption of light in a thin film of arbitrarily low intrinsic optical absorption by embedding it in a judiciously structured photonic environment [12,13]. Indeed, spectrally flat 100% absorption has been recently demonstrated over a full octave of bandwidth in a 2-μm-thick polycrystalline silicon film incorporated into a carefully designed planar cavity [14]. However, like most coherent-enhancement schemes, CPA is achieved only at discrete resonant wavelengths [11,14-16]. Unavoidably, resonant enhancements within an optical cavity are harnessed over only narrow bandwidths as a result of the fundamental link between the cavity-photon lifetime and the resonant linewidth [17].

Here we demonstrate that an omni-resonance scheme [18,19] can extend CPA continuously over a broad spectral bandwidth, thereby rendering this coherent-enhancement arrangement relevant for solar conversion applications. By severing the link between the cavity-photon lifetime and the resonant bandwidth, omni-resonance leads to the emergence of 'achromatic' resonances that deliver coherently enhanced broadband absorption. We realize this effect in a hydrogenated amorphous silicon (a-Si:H) PIN-diode solar cell incorporated into an unpatterned asymmetric planar Fabry-Pérot cavity produced via standard deposition technology, and is thus scalable in principle to large areas. Realizing omni-resonance does not necessitate modifying the cavity structure as in previous work on 'white-light' cavities [20-25]. Instead, incident light is preconditioned using an alignment-free optical arrangement that assigns to each wavelength an appropriate angle of incidence [18,19]. In effect, we introduce into the external field angular dispersion that is equal in magnitude but opposite in sign to that of the cavity. Consequently, angular dispersion cancellation allows all wavelengths within a selected band – that may extend over multiple free spectral ranges of the cavity – to resonate simultaneously, resulting in continuous-wavelength CPA and a concomitant boost in the external quantum efficiency (EQE). The combination of CPA and omni-resonance leads to a doubling in the photocurrent harvested in the near-infrared spectral range 660–740 nm at the a-Si:H band edge. Although our proof-of-principle demonstration makes use of a thin-film a-Si:H solar cell, the combination of CPA and omni-resonance is materials-agnostic and can be exploited with any other form of absorbing photonic structure, such as organic photovoltaics, two-dimensional materials, and perovskites. In all such cases, complete broadband optical absorption is ensured in the solar cell regardless of its thickness or its intrinsic absorption. This result therefore fulfills a long-standing goal in optical physics whereby the inherent advantages of coherently enhanced optical effects – normally restricted to discrete wavelengths [26,27] – is harnessed over a broad continuous spectrum.

# Results

**Overall device architecture**

The solar device illustrated in Fig. 1a brings together CPA and omni-resonance to enable broadband coherently enhanced absorption in a thin-film solar cell. At the heart of this device is a thin-film a-Si:H

PIN-diode solar cell provided with two transparent conductive contacts and sandwiched between (1) a back-reflector formed of a dielectric Bragg mirror providing broad spectral reflection maintained over a large span of incidence angles, and (2) a dielectric spacer and a partially reflective dielectric front Bragg mirror. An example of a fabricated device is shown in Fig. 1b. The back-reflector and front mirror define an asymmetric planar Fabry-Pérot cavity designed to realize the conditions for CPA [12]. However, because the complete absorption associated with CPA is achieved in the solar cell only on resonance (all other wavelengths reflect back from the cavity), we exploit the phenomenon of omni-resonance [18] to extend CPA to a broad *continuous* spectral band. Making use of a light preconditioning system that endows broadband radiation incident on the CPA cavity with judicious angular dispersion, wavelengths across a continuous omni-resonant bandwidth are fully absorbed, thereby enhancing the EQE of the solar cell and boosting the photocurrent. The omni-resonant bandwidth is *not* related to the photon lifetime and can extend far beyond the cavity linewidth. We have designed the system described here to increase the near-infrared EQE in the vicinity of the electronic bandgap edge of a-Si:H in the spectral range 660 – 740 nm, but the same strategy can be exploited with other material systems and in any spectral band.

**Theoretical foundations for the device functionality**

We start by noting that an optically thin layer having low absorption $\mathcal{A} \ll 1$ can always be made to absorb at least 50% of the incident light by engineering its photonic environment [28-30], a configuration also known as a Salisbury screen [31,32]. In the case of a layer of non-negligible thickness, a maximal absorption of $\mathcal{A}_{\text{tot}} = \frac{1}{8}\{2 + \bar{\mathcal{A}} + \bar{\mathcal{A}}^{-1}\}$, is achieved in a *symmetric* cavity ($\frac{1}{2} \leq \mathcal{A}_{\text{tot}} \leq \frac{2}{3}$) when both mirrors have a reflectivity $R = \frac{3\bar{\mathcal{A}}-1}{\bar{\mathcal{A}}(3-\bar{\mathcal{A}})}$, where $\bar{\mathcal{A}} = 1 - \mathcal{A}$ [12]. CPA allows for $\mathcal{A}_{\text{tot}}$ to be raised unconditionally to 100% (independently of $\mathcal{A}$) by arranging for the field to impinge on *both* sides of the symmetric device after setting the mirrors' reflectivity to $R = \bar{\mathcal{A}}$ and – critically – adjusting the relative phase and amplitude of the two fields [10,28,33]. This interferometric configuration is *not* suitable for solar applications where it is prohibitively difficult to maintain a spectrally varying relative phase between two beams of sunlight. However, complete absorption ($\mathcal{A}_{\text{tot}} = 1$) can nevertheless be achieved with a single incident field by utilizing an *asymmetric* structure comprising of a back-reflector and a partially reflective front mirror having $R = \bar{\mathcal{A}}^2$ [14] (Fig. 2a). This configuration is reminiscent of critical coupling that occurs when radiation is coupled into a cavity at a rate that matches its dissipation rate [34,35]. Although critical coupling is usually studied in micro-resonators, it is also realizable in planar structures [36].

The above description of CPA applies only to the resonant free-space wavelengths $\lambda_m = 2nd/m$ at normal incidence, where $m$ is the resonance order, $d$ is the cavity thickness, and $n$ is its refractive index; Fig. 2a. Omni-resonance then enables extending the spectral range of CPA over a *continuous* band (Fig. 2b), much wider than the resonant linewidth, which is indispensable for solar-energy applications. The underlying principle can be understood by first noting that tilting the incident radiation by an angle $\theta$ with respect to the cavity normal *blue*-shifts the resonances $\lambda_m(\theta) < \lambda_m(0)$. Indeed, *any* free-space wavelength $\lambda$ can resonate with the $m^{\text{th}}$ cavity mode if it is assigned an external incidence angle $\theta(\lambda)$ satisfying the omni-resonance condition $\sin^2(\theta(\lambda)) = n^2\{1 - (m\lambda/\lambda_0)^2\}$, where $\lambda_0 = 2nd$ is the free-space wavelength of the fundamental resonant mode at normal incidence [18]. Maintaining the omni-resonant condition over the selected spectral band therefore requires preconditioning the incident light such that each $\lambda$ is assigned to the appropriate incidence angle $\theta(\lambda)$. Consequently, optical absorption in a solar cell embedded in such a cavity is resonantly enhanced over the omni-resonant span, thereby boosting the harvested photocurrent over bandwidths relevant to solar-energy.

To gain insight into the construction of the required light preconditioning system, we plot in Fig. 2c the angular-spectral trajectories of the cavity resonances for different modes $m$ and refractive indices $n$;

the impact of the cavity thickness $d$ is implicit in the normalization wavelength $\lambda_0$. To precompensate for the angular dispersion of each slanted resonance-curve that is intrinsic to planar cavities, we render the incidence angle wavelength-dependent, to first-order $\theta(\lambda) = \psi - \beta(\lambda - \lambda_m)$ for the $m^{\text{th}}$ resonance; here $\psi$ is the cavity tilt angle with respect to normal and $\beta$ is the angular dispersion introduced, which must be equal in magnitude but opposite in sign to the angular dispersion of the cavity resonance. By virtue of the angular dispersion introduced into the incident light, the system 'de-slants' the spectral trajectory of the selected mode (Fig. 2d), and a broad omni-resonant spectrum associated with a single underlying resonant mode is consequently made available. Note that 'anomalous' angular dispersion is required: shorter wavelengths must impinge upon the cavity at larger angles. Furthermore, reducing the required $\beta$ necessitates reducing $n$, which we achieve by adding a dielectric spacer of lower index than Si and larger thickness than the PIN-diode, which incidentally reduces the cavity free spectral range.

**Fabrication of the resonant solar device**

The question remains concerning the feasibility of manufacturing such a resonant structure and the impact of fabrication tolerances on its performance. The fabrication steps are outlined in Fig. 3 in which only standard planar deposition technologies are utilized. After cleaning a square glass substrate (25×25 mm$^2$ area, 1-mm-thick; step 1), a multilayer dielectric Bragg back-reflector is deposited via e-beam evaporation at 200 °C (step 2) in lieu of the more traditional metal back-reflector. To ensure that reflection is maintained over the spectral window of interest even at large angles of incidence, the back-reflector is formed of two Bragg mirrors (Methods), whose dual-band design ensures full reflectivity for light incident between 0° and 55° across the spectral band from 600 nm to 850 nm (Supplementary Information).

A transparent conductive oxide layer of aluminum-doped zinc oxide (AZO) is deposited via atomic-layer deposition (ALD) at 150 °C as a back contact (step 3). The PIN-junction was produced using silane gas (SiH$_4$) and H$_2$ to deposit reduced a-Si:H, and dope the layers accordingly, via plasma-enhanced chemical vapor deposition (PECVD operating at 13.56 MHz; step 4). To enable contacting the bottom AZO layer, we lithographically etch the corners of the a:Si-H layers to expose the AZO (step 5) before depositing an AZO layer via ALD at 130 °C as a top contact (step 6), followed by sputtering a thin gold layer (step 7). We lithographically define an outer ring of gold for contact with the bottom AZO layer and a central feature in the form of a half-circle for contacting the top AZO layer (step 8). The top and bottom AZO layers are joined and envelope the PIN-diode. However, because the vertical dimensions of the layers are orders-of-magnitude smaller than the transverse dimensions, etching a trench into the top AZO layer separates the top and bottom contacts and defines the active area (step 9). At this stage we have a fully functional solar cell atop a back-reflector. In a final step, we deposit a silica spacer (of thickness 2, 4, or 10 μm) and a second dielectric Bragg mirror on the solar-cell active area (step 10). Because the optical absorption $\mathcal{A}$ in the cavity is wavelength-dependent, the required mirror reflectivity must in turn vary with wavelength $R = R(\lambda) = [\bar{\mathcal{A}}(\lambda)]^2$, which necessitates an aperiodic multilayer structure [14]. Here we make use of a partially reflective dielectric Bragg mirror that provides a best fit and comprises 3 bilayers of SiO$_2$ and TiO$_2$. See Supporting Information for details of the fabrication process and optical characterization of the individual layers. The thickness of the PIN-diode is ∼ 360 nm, each contact ∼ 300 nm, the back-reflector ∼ 5.7 μm, and front mirror ∼ 612 nm. The device thickness is therefore dominated by the spacer and back-reflector.

**Modeling and optical characterization**

To establish the operating baseline for absorption and EQE, we first characterize a bare solar cell (Fig. 4a) fabricated by following the steps illustrated in Fig. 3 but eliminating the back-reflector, the spacer, and the front mirror. The measured absorption and EQE are provided in Fig. 4b,c (Methods); the measured short-circuit current density under the solar spectrum irradiance AM 1.5G is ≈ 5 mA/cm$^2$. We are interested in the near-infrared spectral regime 660 – 740 nm where optical absorption and consequently the EQE drop rapidly (the short-circuit current density produced in this spectral window is $\approx 0.26$ mA/cm$^2$). The full CPA-enhanced solar device (Fig. 4d) is a resonant structure in which the back-reflector and front mirror sandwich the solar cell, as is clear from the measurements shown in Fig. 4e,f (for a spacer thickness of 2 μm). Increasing the spacer thickness of course reduces the free spectral range and thus increases the number of resonances in the spectral window of interest.

To assess the angular dispersion required to achieve omni-resonance, we trace the resonant trajectories in the angular-spectral domain; i.e., the absorption spectrum for all incidence angles $\mathcal{A}_{\text{tot}}(\lambda, \theta)$. The calculated and measured absorption spectra are plotted in Fig. 5a and Fig. 5b, respectively. The calculations (Fig. 5a) model the multilayered CPA solar device using the transfer-matrix method employing measured wavelength-dependent optical constants for the individual layers, which are obtained by spectroscopic ellipsometry (Supplementary Information). The measurements (Fig.5b) were carried out using the setup illustrated in Fig. 5f, where light from a halogen lamp is collimated and directed to the active area of the CPA solar device and the reflected $R_{\text{tot}}$ and transmitted $T_{\text{tot}}$ light are collected, from which we have $\mathcal{A}_{\text{tot}} = 1 - R_{\text{tot}} - T_{\text{tot}}$ (Methods). We note the excellent agreement between the calculated and measured spectral trajectories of the resonances and thence the associated angular dispersion $\beta$. Moreover, as predicted, increasing the thickness of the dielectric spacer reduces the slope of the spectral trajectories $\beta$ of the resonances and also reduces the free spectral range.

To achieve omni-resonance, we make use of the measured value of $\beta$ for a selected resonance, and design a preconditioning system that introduces angular dispersion $-\beta$ into the field. However, no known optical component can endow broadband light with the desiderata for omni-resonance: angular dispersion that is anomalous *and* has a large magnitude. A grating typically introduces *normal* angular dispersion whose magnitude is limited by the ruling density. Nevertheless, tailoring the geometry by tilting the grating with respect to the incident radiation and tilting the cavity with respect to the grating can switch the angular dispersion from normal to anomalous, and a lens can subsequently increase the magnitude of $\beta$ [18]. We construct such a system with $\lambda_m \approx 710$ nm to guarantee that each wavelength $\lambda$ is incident on the cavity at the angle $\theta(\lambda)$ satisfying the omni-resonance condition for specific values of the cavity tilt angle $\psi$ (Fig. 5g). The calculated absorption spectrum $\mathcal{A}_{\text{tot}}(\lambda, \psi)$ shown in Fig. 5c,d after introducing the preconditioning system depicted in Fig. 5g reveals that the trajectories of the resonances are deformed to yield a large bandwidth over which they are 'de-slanted'. The measured absorption spectrum $\mathcal{A}_{\text{tot}}(\lambda, \psi)$ after implementing the light-preconditioning system in the path of incident light is shown in Fig. 5e, and good agreement with the theoretical predictions in Fig. 5d is clear. Note that the continuous resonant bandwidth is *not* a result of a large number of resonances merging or overlapping, but instead each de-slanted 'achromatic' resonance now provides a continuous omni-resonant bandwidth independently of the linewidth of the bare cavity.

**Omni-resonant CPA-enhancement of the EQE**

We now proceed to demonstrate that the CPA-enhanced solar device does indeed improve the EQE over the omni-resonant bandwidth through spectrally *and* angularly resolved EQE measurements. First, using the setup shown in Fig. 6d we direct spectrally resolved light from a monochromator to CPA-enhanced solar device (in absence of a light preconditioning system) at an angle of incidence $\theta$. By scanning $\theta$ from 0° to 80° and scanning the wavelength from 500 nm to 800 nm (Methods), we obtain the spectrally and angularly resolved EQE plotted in Fig. 6a for the three spacer thicknesses. The measurements follow closely the corresponding absorption results (Fig. 5a,b) except for the decay in EQE once the bandgap edge of a-Si:H is reached at ~ 730 nm. By introducing the light preconditioning system (diffraction grating plus lens) between the monochromator and the solar device (Fig. 6e), we carry out the spectrally resolved measurements of the EQE in the omni-resonant configuration while varying the cavity tilt angle $\psi$. The measurements (Fig. 6b) are in good agreement with the corresponding optical absorption (Fig. 5c-e) except for the limiting effects of the a-Si:H band edge; for example, whereupon 90% is absorbed at the 775-nm resonance (Fig. 4e), the EQE is enhanced by only 1% (Fig. 4f).

To highlight the improvement in EQE as a result of the omni-resonant CPA-enhanced solar device, we compare in Fig. 6c the EQE of the bare solar cell at normal incidence over the spectral range of interest (Fig. 4c) to the omni-resonant result corresponding to $\psi = 50°$, $44°$, and $44°$ for the spacer thicknesses 2, 4, and 10 μm, respectively, corresponding to specific achromatic resonances. The omni-resonant EQE is consistently higher in the targeted wavelength span of 660 – 740 nm. Table 1 summarizes the improvement in the omni-resonantly generated photocurrent of solar cells shown in Fig. 6c for the three dielectric spacer thicknesses. The short-circuit photocurrent density is $J_{sc}(\psi_o) = q \int d\lambda\, \eta(\lambda, \psi_o)\, \varphi(\lambda)$ where $q$ is the electron charge, $\varphi(\lambda)$ is the photon flux of the solar spectrum irradiance AM 1.5G, $\eta(\lambda, \psi_o)$ is the EQE at a fixed cavity tilt angle $\psi_o$ that optimally achieves omni-resonance, and we have carried out the spectral integral over two ranges $\Delta_1 = 660{:}700$ nm (40 nm) and $\Delta_2 = 660{:}740$ nm (80 nm). We find that several configurations (2-μm and 4-μm spacer thicknesses) lead to more than a doubling of $J_{sc}$ over the $\Delta_1$ range and slightly less than a doubling in the $\Delta_2$ range.

**Planar light preconditioning configuration**

This proof-of-principle demonstration validates the feasibility of omni-resonant spectral broadening of CPA beyond the cavity resonance linewidth. However, the particular realization of light preconditioning utilized (diffraction grating plus lens [18,19]; Fig. 7a) is not suitable in the context of harvesting solar-energy. The system is bulky and cannot be deployed over large areas. Critically, the lens renders the system shift-*variant*, and the performance is thus sensitive to the position *and* size of the incident field. This is clear in the measurements of angular dispersion produced by the system shown in Fig. 7a for two different beam sizes at the grating (1 mm in Fig. 7e and 10 mm in Fig. 7i), where increasing the field size blurs the imparted angular dispersion (Methods). An ideal light preconditioning system produces the requisite angular dispersion in a flat, thin, shift-*in*variant structure that is insensitive to the incident field size and position. In other words, this is an alignment-free system in which the incident field can impinge on any part of the entrance, and the three layers (the two gratings and the prism) do not require specific relative transverse positioning with respect to each other.

Because the role of the lens is to 'amplify' the grating's angular dispersion, one may replace the lens with an appropriately tilted second grating (Fig. 7b). Measurements plotted in Fig. 7f confirm that angular dispersion similar to that in Fig. 7e is produced, and shift-invariance is now realized (Fig. 7j). Although only planar devices (gratings) are used, this alignment-free system remains bulky because of the required angular tilt between the two gratings. The relative tilt between the two gratings can be replaced with a dielectric prism (3D-printed from VeroClear polymer of a refraction index 1.47 and prism angle 41°) that deflects the diffracted field from the first grating appropriately, as shown in Fig. 7c, while maintaining the angular dispersion and shift invariance (compare Figs. 7g,k to Figs. 7f,j). A final design that satisfies the above-listed requirements replaces the macroprism in Fig. 7c with a one-dimensional microprism array (0.5-mm-thick, 127-μm-wide microprisms, molded PMMA polymer of refraction index 1.49 and microprism angle 41°; PR 712 from Orafol Fresnel Optics GmbH) as shown in Fig. 7d and Supplementary Fig S9. Here the three component layers (two gratings and thin sheet of microprisms) can lie atop of each other with no need for any particular separation between them, resulting in a substantially reduced-thickness arrangement. The measurement results obtained with this alignment-free configuration containing the *micro*prism array shown in Fig. 7h,l are in full correspondence with the results from a single *macro*prism in Fig. 7g,k. See Methods and Supplementary Information for details of the experimental configurations used.

## Discussion

Our approach offers a new avenue for improving absorption in solar cells that is distinct from light trapping via surface patterning. Fully deterministic coherent enhancement is harnessed in our strategy for incoherent radiation without the traditional drawback of narrow resonant bandwidths. The planar structure we designed and realized exploits a variant of CPA to maximize optical absorption in a multi-layer thin-film structured cavity configuration rather than the previously studied single-absorbing-layer arrangements [11,12,14]. Therefore, using purely optical methods that are independent of the specific material system utilized, and without changing the structure of the thin-film solar cell itself, we boost the near-infrared EQE. This capability, which has been demonstrated to date only at specific resonant wavelengths, is extended here via omni-resonance over a continuous targeted bandwidth of 80 nm using a single incoherent optical field in a non-interferometric scheme. This is a crucial step towards utilizing CPA and other resonantly enhanced effects – inherently restricted to discrete wavelengths – in practical energy-harvesting technologies by delivering coherent enhancements over a broad bandwidth. Future research will focus on the fundamental question of the limit of omni-resonant-based enhancement of CPA that can impact the power conversion efficiency of a solar cell. Although CPA guarantees that light is fully absorbed in the solar cell independently of the intrinsic absorption of the material or its thickness, not all the absorbed photons will yield free carriers that contribute to the usable electrical energy. Note that our proof-of-principle system (particularly the back-reflector) was designed to optimize optical absorption in the near-infrared over the spectral range 660 – 740 nm. Future work will be directed along two paths. First, we will investigate the utility of the combination of CPA and omni-resonance over the visible and near-infrared, rather than only the near-infrared as demonstrated here. Second, as a potential route to transparent solar cells, we will explore thinner solar cells with even less absorption in the visible – thus rendering the cell more transparent – and determine the maximum possible boost in the near-infrared photocurrent.

It remains an open question whether a *single* textured surface (e.g., a metasurface [37,38]) can provide the requisite angular dispersion to achieve omni-resonance in lieu of the multi-surface systems shown in Fig. 7. Recent progress has been achieved in realizing one of the above-mentioned desiderata, namely producing anomalous diffraction from a single surface [39-41]. It remains to combine this anomalous diffraction with a large magnitude of the accompanying angular dispersion.

Relying on a spectrally selective all-dielectric back-reflector paves the way to thin-film solar-cell devices that are almost transparent except for a specific spectral band (e.g., the infrared) in which optical absorption is maximized. This may establish the viability of transparent solar cells in building-integrated photovoltaics and automotive applications, for instance [9]. These applications will benefit from developing a metasurface realization of the light preconditioning system for omni-resonance that further reduces the thickness of the microprism-based system shown in Fig. 7d.

We note that the underlying principle of omni-resonance is in the exploitation of correlations introduced into the spatio-temporal spectrum (angle-wavelength) of the incident radiation. This approach therefore falls under the rubric of space-time wave packets [42], which are propagation-invariant (diffraction-free and dispersion-free) coherent [43,44] or incoherent [45] fields whose unique characteristics stem from introducing angular dispersion into their spatio-temporal spectrum to compensate for the angular dispersion intrinsic to free propagation. In our work here, the angular dispersion introduced into the field compensates for that intrinsic to resonant cavity modes. More generally, both omni-resonance and space-time wave packets are ultimately examples of the utility that can be harnessed by 'entangling' different degrees of freedom of the optical field [46,47] rather than remaining independent of each other.

**Table 1 | Integration of EQE over spectral range of interest**

|  |  | Bare solar cell | CPA-enhanced solar cell | | |
|---|---|---|---|---|---|
|  |  |  | 2 µm | 4 µm | 10 µm |
| $\Delta_1$ = 40 nm | $J_{sc}$ (mA/cm$^2$) | 0.184 | 0.395 | 0.401 | 0.329 |
| $\Delta_1$ = 40 nm | % photocurrent | 100% | 212% | 218% | 179% |
| $\Delta_2$ = 80 nm | $J_{sc}$ (mA/cm$^2$) | 0.258 | 0.497 | 0.495 | 0.405 |
| $\Delta_2$ = 80 nm | % photocurrent | 100% | 191% | 192% | 157% |

# Materials and methods

**Structure of the SiO$_2$/TiO$_2$ Bragg reflector.** The back-reflector is a dual band mirror comprising a sequence of two Bragg mirrors, each consisting of 13 bilayers of SiO$_2$ (L: low refractive index) and TiO$_2$ (H: high refractive index). Starting from the substrate, the first Bragg mirror has 13×[115.8 nm (L), 76.2 nm (H)] for lower wavelengths, and 115.8 nm (L), 13×[92.9 nm (H), 141.3 nm (L)] for the higher wavelengths.

**Angular and spectrally resolved optical absorption measurements of the CPA-enhanced solar device.** Measuring the angular-resolved spectral absorption $\mathcal{A}(\lambda, \theta)$ is key to estimating the required angular dispersion for omni-resonance. To measure $\mathcal{A}(\lambda, \theta)$, we measure the normalized transmission $T_{\text{tot}}$ and reflection $R_{\text{tot}}$, from which we have $\mathcal{A}(\lambda, \theta) = 1 - T_{\text{tot}} - R_{\text{tot}}$, as given in Fig. 5b. To measure $T_{\text{tot}}$ and $R_{\text{tot}}$, we use a 2-mm-diameter unpolarized, spatially filtered beam of broadband white light from a quartz tungsten-halogen lamp (Thorlabs QTH10, 50 mW power) incident at different angles onto the device in the spectral range 500 – 800 nm. The transmitted and reflected light beams are coupled into a multi-mode fiber (Thorlabs M69L02; core diameter of 300 µm, 0.39 NA) that delivers the light to a spectrometer (Jaz, Ocean Optics). These measurements are normalized with respect to the spectral density of the initial beam in absence of the sample.

**Omni-resonance optical absorption measurement.** The required angular dispersion for the 2, 4, and 10-µm spacer thicknesses are 0.41, 0.35, and 0.31 °/nm respectively. To introduce this angular dispersion into the incident radiation, we use a diffraction grating (Thorlabs GR25-1208; 1200 lines/mm, area 25×25 mm$^2$) that introduces an angular dispersion of ~ 0.07 °/nm for an angle of incidence of 44° and then 'magnify' this angular dispersion via a spherical lens. The required magnification factor $M$ is 5.9, 5.1 and 4.3 for the 2, 4, and 10-µm spacer thicknesses, respectively. If the distance from the grating to the lens is $d_1$ and the lens focal length is $f$, then a magnification factor of $M$ is realized when $d_1 = (M + 1)f$. We thus implement distances $d_1$ of 172.5, 152.5 and 132.5 mm for the 2, 4, and 10-µm spacer thicknesses, respectively. The sample is placed in the image plane, at a distance $d_2 = d_1/M$ from the lens. Because omni-resonance occurs at particular values of the cavity tilt angle $\psi$, we mount the sample on a rotational stage (Thorlabs MSRP01) to alter $\psi$.

The system is designed such that the wavelength 710 nm lies along the optical axis of the imaging system (so that the angle $\psi$ is measured with respect to it; see Supplementary Information). The lens aperture of 25 mm limits the collected bandwidth of the beam reaching the sample to ~ 660 – 760 nm. The back-reflector eliminates transmission through the device in this spectral window, so that $T_{tot} \approx 0$. By measuring $R_{tot}$ for different tilt angles $\psi$, we obtain the absorption $\mathcal{A}(\lambda, \psi) = 1 - R_{tot}$, as given in Fig. 5e. The reflected light is collected by a multi-mode fiber (Thorlabs M69L02; core diameter of 300 μm, 0.39 NA) that delivers the light to a spectrometer (Jaz, Ocean Optics). The reflected spectrum is normalized in the range 660 – 760 nm with respect to the collected spectrum when the sample is replaced with a back-reflector alone.

**External Quantum efficiency measurements.** The EQE is measured using a commercial tool (model QEX10) designed for photovoltaic measurements. An ellipsoidal reflector focuses light from a Xenon-arc lamp onto the entrance slit of a dual-grating computer-controlled monochromator (Monochromator 22535 NOVRAM) via a mechanical chopper that modulates the light at a 100-Hz frequency and provides a reference signal to the digital lock-in amplifier. The selected wavelength emerges from the monochromator output slit, and bandpass filters attenuate stray and harmonic light. A portion of the monochromatic light is diverted by a beam splitter to a monitoring photodiode via a lens. The light that passes through the beam splitter is focused by a concave mirror onto the device under test. Integral to the measurement setup is a reference detector whose EQE is calibrated against a NIST reference. A typical EQE measurement is carried out in the following order. First, the reference detector is mounted in lieu of the device under test, and the photocurrent measured at each wavelength is used to assign an EQE value to subsequently measured photocurrents by the device at the same wavelength. Second, the calibration detector is replaced by the device under test and the photocurrent is recorded at each wavelength, and the EQE is estimated in reference to the photocurrent measured in the first (calibration) step. The measured current from the monitoring photodiode is used to compensate for any drift arising from the fluctuations of the lamp's power.

The angular-resolved spectral EQE measurements reported in Fig. 6a are performed with the measuring tool in the configuration shown in Fig. 6d. The measurements are taken over the wavelength range 500 – 800 nm (1-nm resolution) for incident angles from 0° to 80° in steps of 2°. The measured EQE resonances are broader than those recorded in the angular-resolved spectral absorption shown in Fig. 5b. We believe that this discrepancy occurs because the monochromatic light beam from the monochromator is not perfectly collimated.

The omni-resonance EQE measurements reported in Fig. 6b are performed with the measuring tool after modifying the configuration to that shown in Fig. 6e after collimating the beam from the monochromator using a concave mirror, a spherical lens (80-mm focal length), and an aperture to spatially filter the 3-mm-diameter beam. The measurements are taken over the wavelength range of 660 – 760 nm (2-nm-resolution) for different tilt angles $\psi$ ranging from 0° to 60° in steps of 2°. Once again, the measured EQE resonances are broader than those recorded in the omni-resonant optical absorption shown in Fig. 5d,e. We also believe that this discrepancy occurs because the monochromatic light beam from the monochromator is not perfectly collimated.

**Angular-dispersion measurement.** We have measured the angular dispersion produced by four different light preconditioning configurations:

(1) *Grating-lens configuration*: This is the system described above where a diffraction grating is placed a distance $d_1$ from a spherical lens of focal length $f$ (Fig. 7a). The configuration was designed to produce an angular dispersion of 0.31°/nm as required for the spacer thickness of 10 μm, as shown in Fig. 7e,i.

(2) *A pair of tilted gratings*: The two identical transmissive gratings have 1400 lines/mm and an area of 24×24 mm² (LightSmyth T-1400-800s Series, polarization-independent, 95% maximum diffraction efficiency in the wavelength range of interest), are tilted by an angle 48.7° with respect to each other, and light is incident on the first grating at an angle 23.4° (Fig. 7b) to produce an angular dispersion 0.31 °/nm (Fig. 7f,j).

(3) *A combination of a pair of parallel gratings and a prism*: The same pair of transmissive gratings from the previous configuration are used with a polymer prism sandwiched between them (Fig. 7c). Light is incident on the first grating at 45°, and the prism angle is 41°.

(4) *A combination of a pair of parallel gratings and a microprism array*: The same pair of transmissive gratings from the previous configuration are used with a PMMA microprism array sandwiched between them (Fig. 7d). Light is incident on the first grating at 45°, and the microprism angle is 41°.

In each configuration, we measure the angular dispersion using a cylindrical lens (250-mm focal length, 60-mm width) in a $2f$ configuration, in which the angle of each wavelength is transformed into a lateral displacement $\ell$ (with respect to the position of the center wavelength of 710 nm) in the lens focal plane. Over the 100-nm bandwidth of interest (660 – 760 nm), the angle $\theta(\lambda)$ associated with a wavelength $\lambda$ is $\theta(\lambda) = \sin^{-1}(\ell/f)$. To measure $\ell$ in the focal plane, a fiber (Thorlabs M69L02; 300-μm-diameter, 0.39 NA) is displaced along a rail in the focal plane. See Supplementary Information for detailed layouts of these experimental arrangements.

**Acknowledgements**

We are grateful to Philippe de Rouffignac, Ed Macomber, Mac Hathaway, and the staff of the Harvard Center for Nanoscale Systems (CNS) for help with the deposition equipment; Jun Cha and Mark Oberdzinski (Optonetic, LLC.; Orlando, Florida) for depositing the dielectric Bragg mirrors; W. Dennis Slafer and B. Diane Martin, Esq. (MicroContinuum, Inc.; Cambridge, Massachusetts) for granting us access to a photospectrometer for EQE measurements; Pieter Kik for spectroscopic ellipsometric measurements; Winston Schoenfeld for solar simulator measurements; Jack Stubbs and Fluvio L. L. Fenoglietto (UCF Institute for Simulations and Training, PD3D; Orlando, Florida) for 3D-printing of the optical prisms and holders; Horst-Peter Mickschat (Orafol Fresnel Optics GmbH, Apolda, Germany) for providing the microprism array; and Joshua Zorn for help preparing the figures. This work was funded by US Air Force Office of Scientific Research (AFOSR) under MURI contract FA9550-14-1-0037 and by the US Office of Naval Research (ONR) under contract N00014-17-1-2458.

**Author contributions**

M.L.V. and A.F.A. conceived and developed the project. M.L.V. and S.S. carried out the simulations. M.B.N., M.L.V., and C.H.V. fabricated the solar cell. A.S., S.S., A.K.J., M.B.N. and C.H.V. carried out the optical absorption and EQE measurements, and processed the data. C.H.V. and A.S. polished the 3D-printed prisms and prepared the figures. All the authors contributed to interpreting the data and writing the manuscript.

Correspondence and requests for materials should be addressed to A.F.A.

**Competing interests**

The authors declare no competing financial interests.

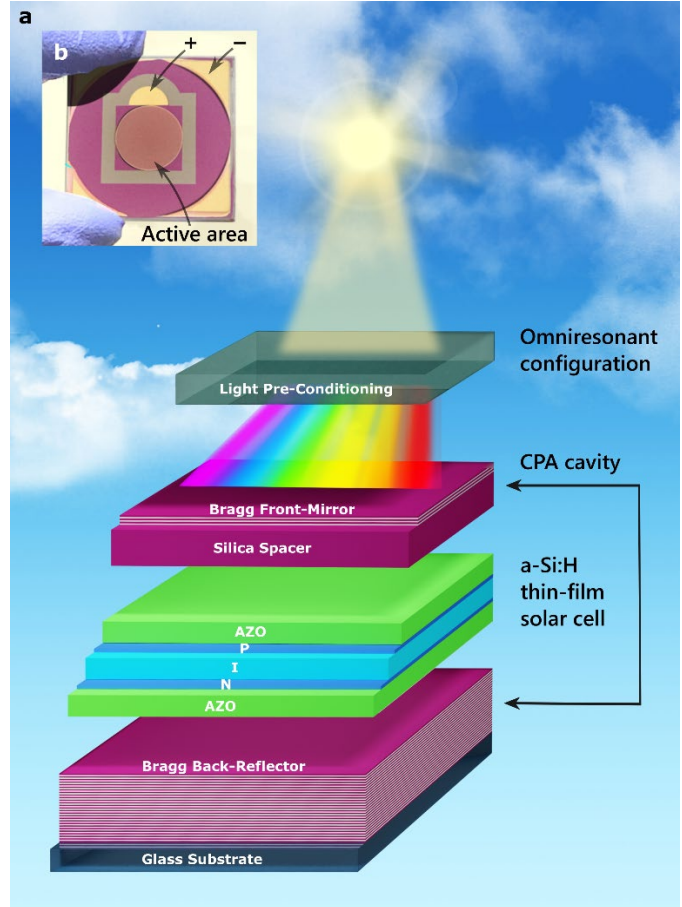

**Figure 1. Architecture of the omni-resonant, CPA-enhanced solar device. a**, Layered structure of the solar cell integrated into a planar Fabry-Pérot cavity with light incident from the top. The layers from bottom-to-top are: glass substrate; broadband dielectric Bragg back-reflector formed of alternating layers of $SiO_2$ and $TiO_2$; transparent, conductive aluminum-doped zinc oxide (AZO) back contact; a-Si:H thin-film PIN-junction solar cell; AZO front contact; silica spacer; and front partially reflective dielectric Bragg mirror. The back-reflector and front mirror constitute a CPA cavity in which the solar cell is integrated. Omni-resonance is achieved through a light preconditioning arrangement that intercepts the incident radiation and assigns to each wavelength a judiciously selected incidence angle to guarantee resonantly enhanced absorption over a continuous spectral span. (b) Photograph of a fabricated CPA-enhanced solar device of area 25×25mm$^2$. The circle in the center is the active area.

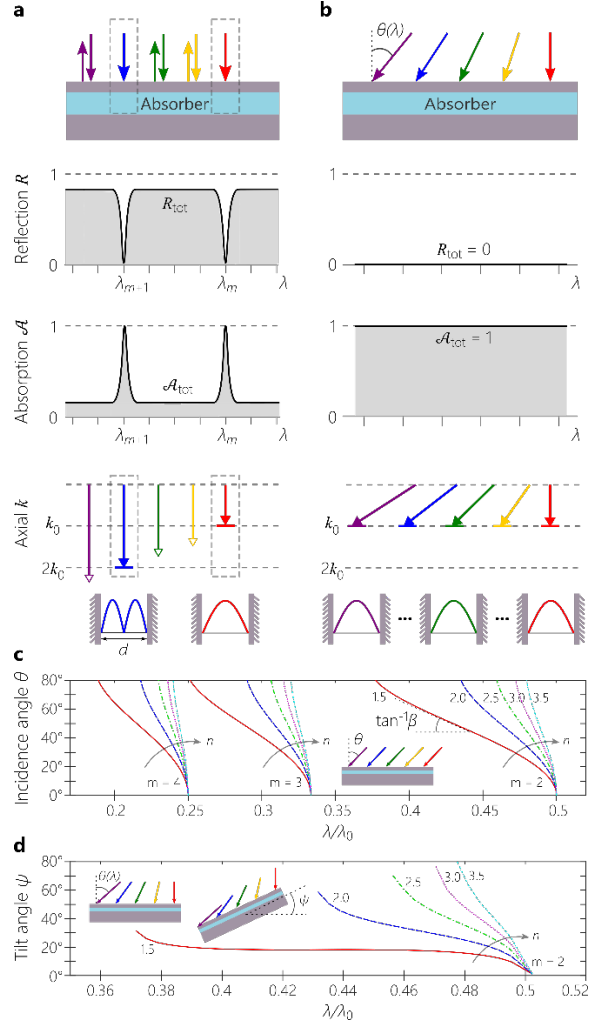

**Figure 2. The physics of optical CPA and omni-resonance. a**, The CPA concept. Broadband collimated light incident on a planar Fabry-Pérot cavity provided with a back-reflector (no optical transmission) and containing a thin layer of low intrinsic absorption $\mathcal{A}$ is reflected back except at the resonant frequencies (where the axial component of the wave vector $k_m$ is an integer multiple of $k_0 = \frac{\pi}{nd}$). If the spectral reflectivity $R$ of the front mirror is designed such that $R = (1 - \mathcal{A})^2$, light at the resonant wavelengths is completely absorbed $\mathcal{A}_{\text{tot}} = 1$ independently of $\mathcal{A}$. **b**, The omni-resonance concept. Incident broadband light is first preconditioned such that each wavelength $\lambda$ is incident at an angle $\theta(\lambda)$ selected to achieve the cavity resonance condition by ensuring that the axial components of the wave vectors over the desired spectrum are all equal to that of one of the bare-cavity resonances. All the wavelengths simultaneously resonate in the *same* longitudinal cavity mode and are fully absorbed, thereby eliminating reflection back from the cavity. **c**, Resonant trajectories for a planar Fabry-Pérot cavity in the angular-wavelength domain for different mode orders $m$ and refractive indices $n$. Here $\beta$ is the angular dispersion of any resonant mode. **d**, The resonant trajectories for $m = 2$ from (**c**) after introducing the omni-resonant preconditioning system that nullifies the cavity angular dispersion $\beta$ and 'de-slants' the resonance in (**c**) to produce an 'achromatic' resonance (the horizontal spectral trajectory).

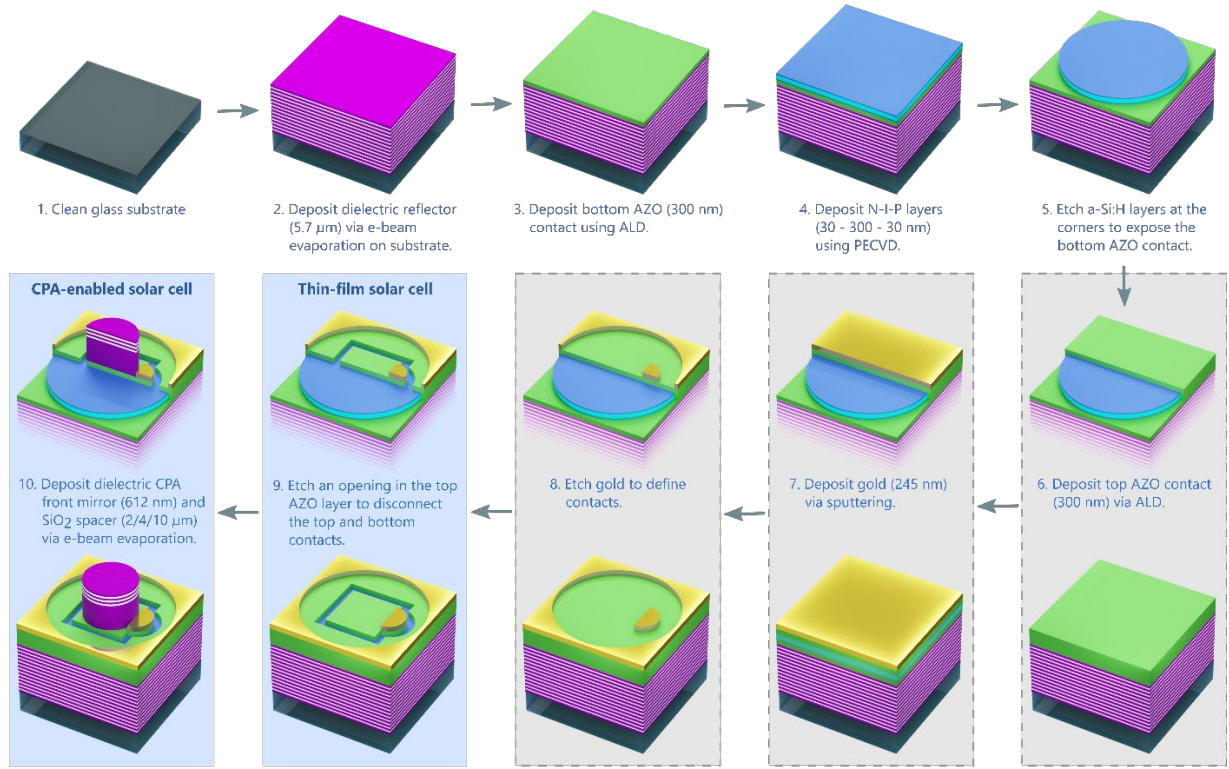

**Figure 3. Fabrication steps of a PIN-diode solar cell integrated into a planar CPA cavity.** In steps 6 through 10 we show the full structure in addition to a cut-away that reveals the relevant internal structure for clarity. All the cells were fabricated using these steps with the only difference being the thickness of the silica spacer deposited in step 10 (thickness of 2, 4, or 10 microns). See Fig. 1b for a photograph of a fabricated device. Note that the vertical scale of the structure (the thickness of the layers; ∼ 7.5 μm in addition to the spacer thickness) is different from the transverse scale (the in-plane feature sizes as shown in Fig. 1b).

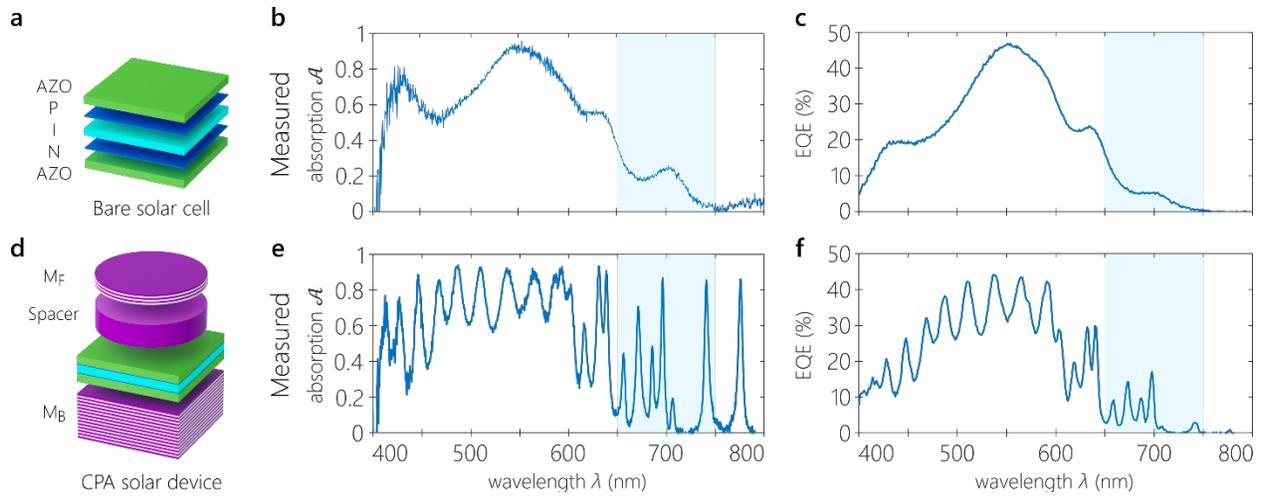

**Figure 4. Optical absorption enhancement in a solar-cell integrated into a CPA cavity. a**, Schematic depiction of the layered structure of a 'bare' solar cell. **b**, Measured optical absorption and (**c**) EQE in the bare solar cell. **d**, Schematic depiction of the layered structure of a solar cell from (**a**) embedded in a CPA cavity. $M_F$: Front, partially reflective Bragg mirror; $M_B$: back Bragg reflector. **e**, Measured optical absorption and (**f**) EQE in the CPA-enhanced solar cell with 2-μm-thick spacer. The blue shading in (**b**,**c**) and (**e**,**f**) highlights the spectral range 660 – 740 nm at the edge of the a-Si:H band edge for which the CPA and omni-resonance systems are designed.

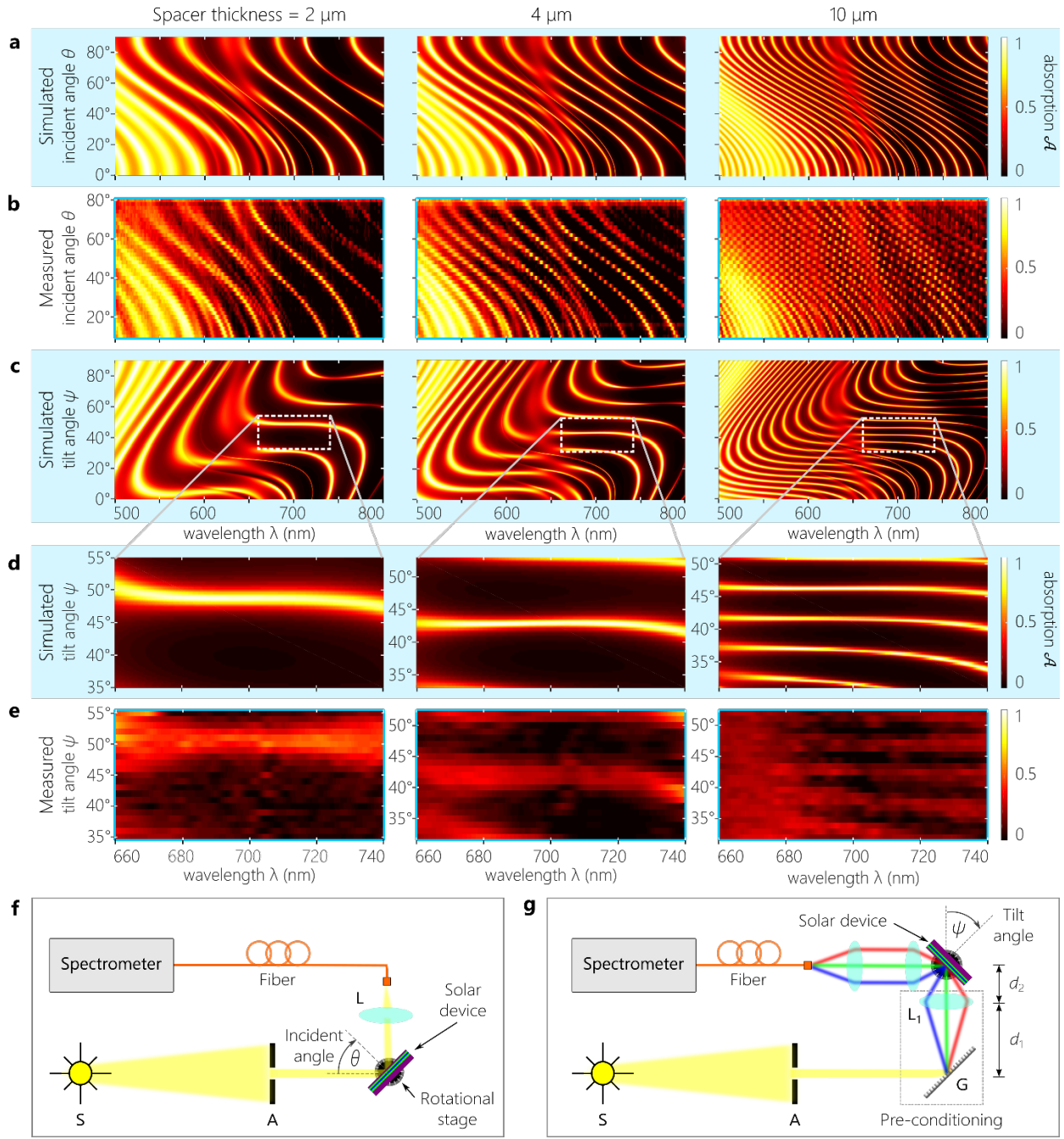

**Figure 5. Optical characterization for solar cavities with different-thickness dielectric spacers**. **a**, Calculated and (**b**) measured spectral absorption in a solar cavity for collimated light at different angles of incidence. **c**, Calculated spectral absorption of an omni-resonant solar cavity at different cavity tilt angles. **d**, Expanded view of (**c**) in the cavity-tilt-angle range of 35°–55° that is accessible in our experimental configuration. **e**, Measured spectral absorption in an omni-resonant solar cavity in the cavity-tilt-angle range of 35°–55°. The shaded panels correspond to simulations. **f**, Schematic of the setup for measuring the optical absorption of the CPA-enhanced solar device and (**g**) after introducing a light-preconditioning system that satisfies the omni-resonance condition. L: Lens; A: aperture; G: grating; S: optical source. See Methods for the distances $d_1$ and $d_2$.

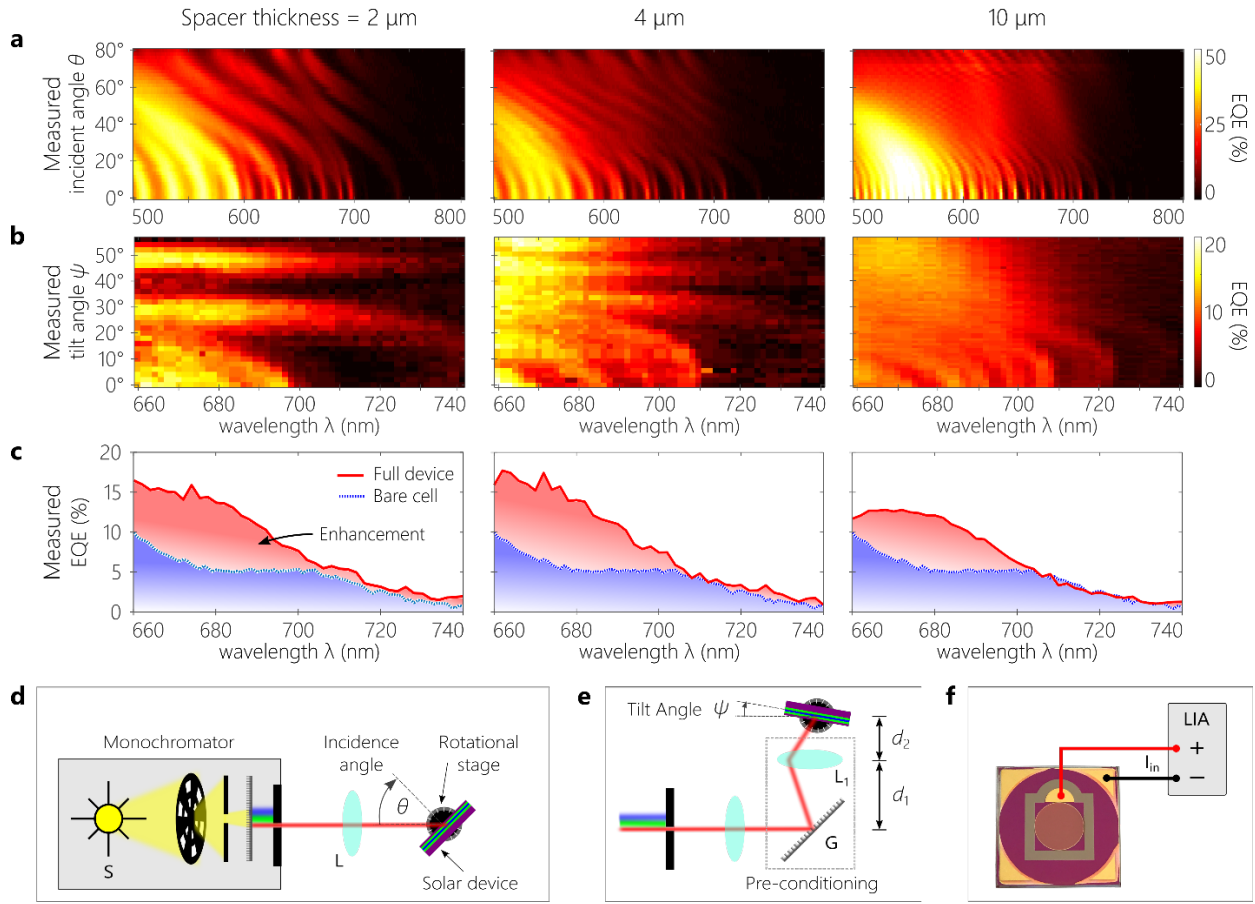

**Figure 6. Characterization of the device quantum efficiency for solar cavities with different-thickness dielectric spacers. a**, Measured spectrally resolved EQE (using collimated incident radiation) with angle of incidence $\theta$ onto a CPA-enhanced solar cavity. **b**, Measured spectrally resolved CPA-enhanced EQE with cavity tilt angle $\psi$ after realization of a light preconditioning system that satisfies the omni-resonance condition. **c**, Measured spectrally resolved EQE across the spectral range 660–740 nm selected from (**b**) at a cavity tilt angle ($\psi = 50°, 44°$ and 44° for spacer thicknesses 2, 4, and 10 µm, respectively) in comparison to the EQE of a reference bare solar cell (from Fig. 4c). **d**, Schematic of the experimental setup for EQE measurements used to obtain the results in (**a**). **e**, Schematic of the experimental setup for the omni-resonant EQE measurements shown in **b** and **c**. **f**, Electrical connections for the fabricated cavity-integrated solar cell. L: Lens; G: grating; S: optical source; LIA: lock-in-amplifier.

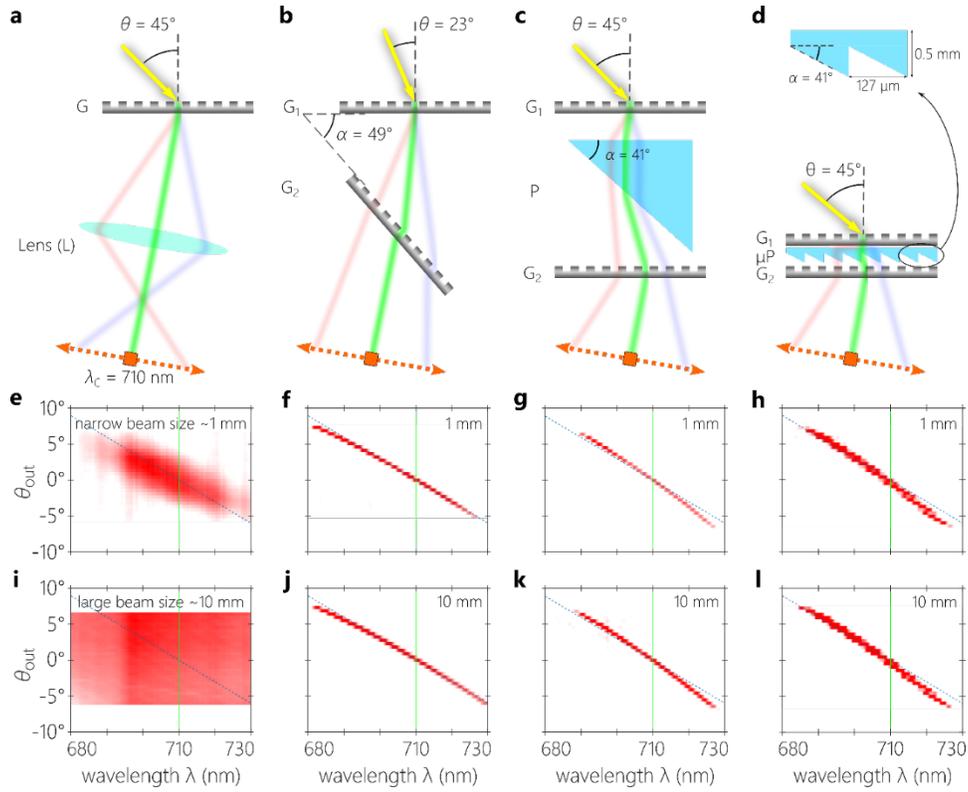

**Figure 7. Preconditioning configurations to achieve omni-resonance. a,** System comprising a reflective diffraction grating (1200 lines/mm) and a spherical lens (Fig. 5g and Fig. 6e). **b**, System comprising two identical transmissive gratings (1400 lines/mm, high-diffraction-efficiency polarization-independent gratings; Methods) tilted with respect to each other. **c**, System comprising two parallel transmissive gratings $G_1$ and $G_2$ (1400 lines/mm) and a prism P for beam deflection. **d**, Same as (**c**) except that the prism P is replaced by a microprism array μP, resulting in a vastly reduced thickness of the optical arrangement (see Supplementary Figure S9). Inset shows a magnified view of the microprisms. **e-h**, Measured angular dispersion of the systems in **a-d** for an incident beam size of ~ 1 mm on the first grating surface (Methods). **i-l**, Measured angular dispersion of the systems in **a-d** for an incident beam size of ~ 10 mm on the first grating surface. Note that the lens used in (**a**) renders the system spatially shift-*variant*, in contrast to the systems in (**b-d**) that are shift-*invariant* and thus independent of incident field size or location. The green vertical lines in (**e-l**) identify the axial wavelength $\lambda_c = 710$ nm.

# Doubling the near-infrared photocurrent in a solar cell via omni-resonant coherent perfect absorption

Massimo L. Villinger[1], Abbas Shiri[1,2], Soroush Shabahang[1,3], Ali K. Jahromi[1], Magued B. Nasr[4], Christopher H. Villinger[1], and Ayman F. Abouraddy[1,5,*]

[1]CREOL, The College of Optics & Photonics, University of Central Florida, Orlando, FL 32816, USA

[2]Department of Electrical and Computer Engineering, University of Central Florida, Orlando, FL 32816, USA

[3]The Wellman Center for Photomedicine at Massachusetts General Hospital, Harvard Medical School, Cambridge, MA 02139, USA

[4]Department of Physics, Boston University, Boston, MA 02215, USA

[5]Department of Materials Science and Engineering, University of Central Florida, Orlando, FL 32816, USA

*corresponding author: raddy@creol.ucf.edu

## Supplementary Information

## Contents

## S.1 Design of the Fabry-Pérot cavity for coherent perfect absorption

The asymmetric Fabry-Pérot cavity designed to satisfy the conditions for one-sided coherent perfect absorption (CPA) consists of a back-reflector $M_B$ and a front mirror $M_F$. The reflectivity of the back mirror is targeted to be unity over the wavelength range of interest to be maintained over a large range of incident angles. The structure of $M_F$ is given in Methods, and simulations of this structure are shown in Fig. S1 confirming that the transmission in the targeted spectral window and angular incidence range of interest is minimal.

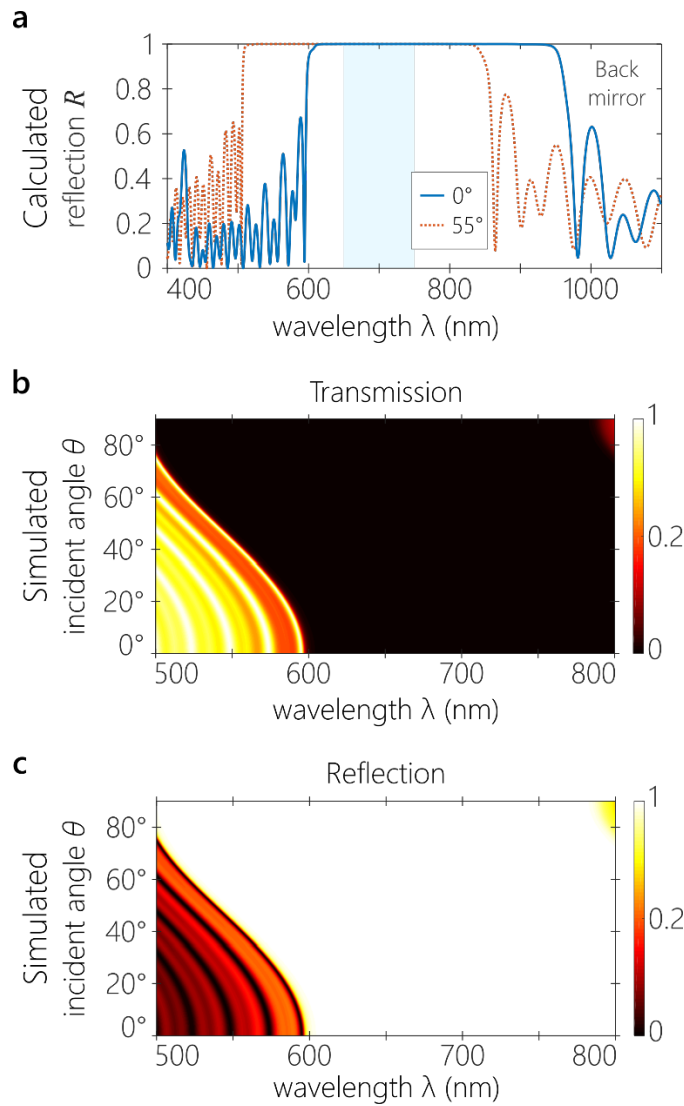

**Figure S1 | Design of the back-reflector in the CPA Fabry-Pérot cavity. a**, Calculated spectral reflection from the back-reflector for light incident from air onto the mirror at normal (0°, blue continuous curve) and oblique (55°, red dotted curve) incidence. **b**, Plot of the calculated angle-resolved spectral transmission through the back-reflector for light incident from air. **c**, Plot of the calculated angle-resolved spectral reflection from the back-reflector for light incident from air. Note that the back-reflector provides full reflectivity (zero transmission) for incidence angles between 0° and 55° in the spectral range from 600 nm to 850 nm.

For such an asymmetric cavity, CPA requires that the reflectivity of the front mirror be

$$R_1 = (1 - \mathcal{A})^2, \qquad (S1)$$

where $\mathcal{A}$ is the single-pass absorption through the cavity. Note that $R_1$ is the internal reflectivity of the front mirror for light incident from within the cavity, which may differ significantly from the reflectivity of the same mirror when light is incident from air. Furthermore, the structure incorporated into the cavity (the PIN-diode with AZO contacts and the silica spacer) is characterized by a wavelength-dependent absorption $\mathcal{A}(\lambda)$, which thus necessitates a wavelength dependent front-mirror reflectivity

$$R_1(\lambda) = \big(1 - \mathcal{A}(\lambda)\big)^2. \qquad (S2)$$

If the absorption in each layer within the cavity is

$$\mathcal{A}_j = 1 - e^{-2k'_j d_j}, \qquad (S3)$$

where $k'_j = 2\pi n'_j / \lambda$ is the extinction coefficient and $n'_j$ is the imaginary part of the refractive index of the $j^{\text{th}}$ layer of thicknesses $d_j$, then the total single-pass absorption resulting from the five layers (AZO-PIN-AZO) is

$$\mathcal{A}(\lambda) = \mathcal{A}_1 + \bar{\mathcal{A}}_1 \mathcal{A}_2 + \bar{\mathcal{A}}_1 \bar{\mathcal{A}}_2 \mathcal{A}_3 + \bar{\mathcal{A}}_1 \bar{\mathcal{A}}_2 \bar{\mathcal{A}}_3 \mathcal{A}_4 + \bar{\mathcal{A}}_1 \bar{\mathcal{A}}_2 \bar{\mathcal{A}}_3 \bar{\mathcal{A}}_4 \mathcal{A}_5, \qquad (S4)$$

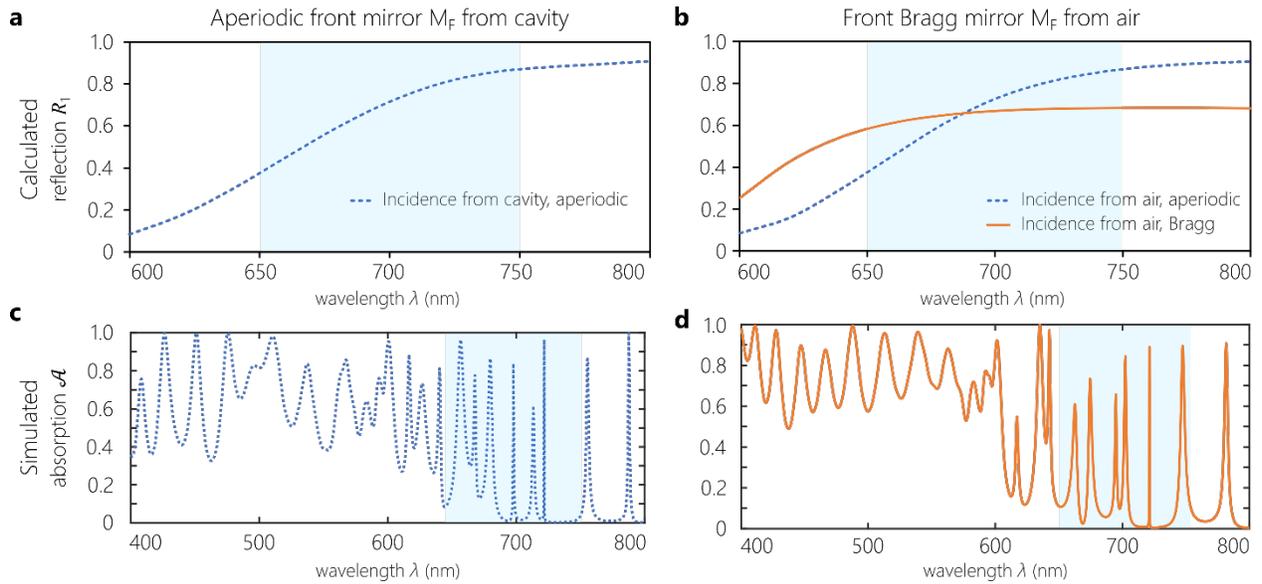

**Figure S2 | Characterization of the front mirror in the CPA Fabry-Pérot cavity. a**, Internal reflectivity of the front mirror $R_1$ for light incident from within the Fabry-Perot cavity onto the mirror. The shaded area corresponds to the spectral range of interest. **b**, Reflectivity of the front mirror for incidence from air when relying on an aperiodic structure that closely flows the theoretical target in a, and that relying on a periodic Bragg structure that approximates the target reflectivity. **c**, Spectral absorption of the CPA Fabry-Pérot cavity when employing the aperiodic front mirror. **d**, Spectral absorption of the CPA Fabry-Pérot cavity when employing a Bragg mirror that approximates the targeted spectral reflectivity of the front mirror.

where $\bar{\mathcal{A}}_j = 1 - \mathcal{A}_j$. Using the measured optical parameters of each layer (see Section S.3 below), we calculate $\mathcal{A}(\lambda)$ and thence the front-mirror reflectivity $R_1(\lambda)$, which we plot in Fig. S2a. The corresponding reflectivity from this mirror for incidence from air is given in Fig. S2b.

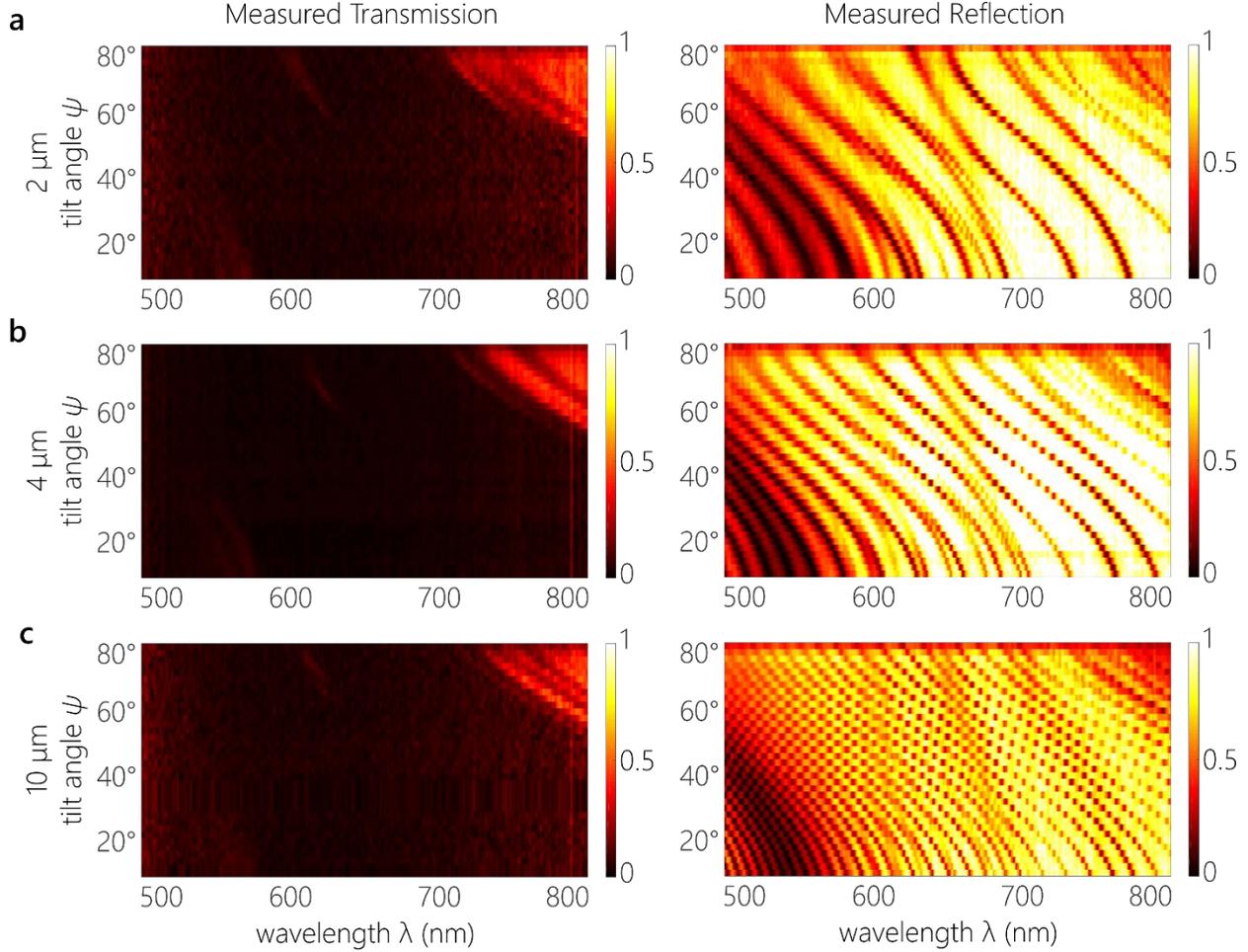

**Figure S3 | Angle-resolved spectral transmission and reflection of the CPA Fabry-Pérot cavity.** Measured spectral transmission $T_{\text{tot}}(\lambda, \theta)$ and reflection $R_{\text{tot}}(\lambda, \theta)$ as a function of the incident angle $\theta$ of collimated light onto the CPA Fabry-Pérot cavity. From this data, the absorption $\mathcal{A}(\lambda, \theta)$ in Fig. 5b of the main text is obtained. **a-c**, Measurements for dielectric spacer thickness (**a**) 2 μm, (**b**) 4 μm, and (**c**) 10 μm.

Constructing a multilayer dielectric mirror requires an aperiodic structure. Indeed, a 7-layer aperiodic mirror consisting of alternating layers of $SiO_2$ and $TiO_2$ can reproduce the targeted reflectivity, validated with the modeling software package FilmStar (FTG Software Associates). Because aperiodic structures place stringent limits on fabrication tolerances, we found the best fit to the targeted reflectivity $R_1(\lambda)$ that is produced by a periodic Bragg mirror structure comprising three periodic bi-layers (Fig. S1b). Such periodic structures have relaxed tolerances with respect to aperiodic designs. To ensure that the approximate periodic structure does not compromise the absorption resonances of the CPA cavity, we compare the absorption spectrum of the cavity calculated using the aperiodic front mirror (Fig. S1c) to the absorption resulting from employing the approximate Bragg mirror (Fig. S1d). We find that the approximate Bragg mirror provides resonances with large absorption peaks that are comparable to those obtained through the use of the exact aperiodic mirror structure.

The absorption $\mathcal{A}(\lambda, \theta)$ plotted in Fig. 5b of the main text is obtained by measuring the spectral transmission and reflection through the cavity as a function of the incident angle of collimated broadband radiation (Methods). The measured angle-resolved spectral transmission and reflection are plotted in Fig. S3 for the three dielectric spacer thicknesses. The measurements confirm that the transmission is zero over the angle range and wavelength range of interest.

## S.2 Solar cell fabrication

We describe here in detail the fabrication steps (depicted in Fig. 3 in the main text) for producing the CPA Fabry-Pérot cavity incorporating the a-Si:H PIN-diode.

**1. Cleaning of the substrates**

The glass substrates (thickness 1 mm) were cleaned by sonication in acetone for 5 min, followed by sonication in methanol for 5 min, dipping in IPA, and rinsing of each substrate individually with more IPA. The substrates are then nitrogen-dried, followed by oxygen-plasma cleaning (40 SCCM of $O_2$ and a 100 W RF power for 4 min). Approximate step-time is 1 h for 8 samples.

**2. Deposition of the back-reflector**

The dual-band multilayer dielectric Bragg back-reflector is deposited via e-beam evaporation at 200 °C (Methods). The substrates with deposited back-reflector were cleaned following the same procedure described in step 1.

**3. Deposition of bottom AZO using an ALD tool**

The bottom transparent conductive oxide (TCO) layer of aluminum-doped zinc oxide (AZO) is deposited at 150 °C. This temperature is lower than traditionally used, and was selected to avoid damaging the dielectric mirror. The growth conditions are as follows:

(i) 39 cycles of ZnO are deposited by the pulsing of $H_2O$ for 0.015 s followed by a 0.015-s pulse of DEZ with 8.5 s of delay in between.

(ii) 1 cycle of $Al_2O_3$ is deposited by the pulsing of $H_2O$ for 0.015 s followed by a 0.015-s pulse of TMA with 8.5 s of delay in between.

The above set of 2 cycles was repeated 45 times. Approximate step-time is 9 h. The samples are then cleaned using the same steps used at the end of step 2 (approximate time is 1 h for 8 samples).

**4. Deposition of N-I-P layers using Plasma-enhanced CVD**

Using $SiH_4$, $H_2$, 1% $PH_3$/Ar as an n-type doping gas, and 10% $B_2H_6/H_2$ as a p-type doping gas in an STS PECVD tool that operates at 13.56 MHz, we deposit the PIN-diode using the following recipe:

Temperature = 200 °C, Pressure = 200 mTorr,

Using $SiH_4$, $H_2$, 1% $PH_3$/Ar as an n-type doping gas, and 10% $B_2H_6/H_2$ as a p-type doping gas in an STS PECVD tool that operates at 13.56 MHz, we deposit the P-I-N junction using the following recipe:

***n-layer*:** Ignition power: 70W, $t$ = 2 s; Deposition power: 20 W, t = 4:20

*Recipe*: $SiH_4$ = 10 SCCM, HF power = 20 W, $H_2$ = 0 SCCM, Temperature = 200 °C, 10% $B_2H_6/H_2$ = 0 SCCM, Pressure = 200 mTorr, 1% $PH_3$/Ar = 20 SCCM

*Expected Thickness:* 30 nm

***i-layer*:** Ignition power: 70 W, $t$ = 2 s; Deposition power: 20 W, t = 38:50

*Recipe:* $SiH_4$ = 58 SCCM, HF power = 20 W, $H_2$ = 20 SCCM, Temperature = 200 °C, 10% $B_2H_6/H_2$ = 0 SCCM, Pressure = 200 mTorr, 1% $PH_3/Ar$ = 0 SCCM

*Expected Thickness:* 300 nm

***p-layer*:** Ignition power: 70 W, $t$ = 2 s; Deposition power: 20 W, t = 1:30

*Recipe:* $SiH_4$ = 40 SCCM, HF power = 20 W, $H_2$ = 0 SCCM, Temperature = 200 °C, 10% $B_2H_6/H_2$ = 20 SCCM, Pressure = 200 mTorr, 1% $PH_3/Ar$ = 0 SCCM

*Expected Thickness:* 30 nm

Approximate step time is 45 m for each run.

## 5. Photolithography to expose the bottom AZO contact

In this step, we etch away the a-Si:H layer at the corners of the substrate to expose the bottom AZO conductor that will be used subsequently as a bottom contact.

(i) Spin-coat S1818 Shipley photoresist for 5 s at 500 RPM with a 100 RPM/sec ramp, followed by a 45-s spin at 2000 RPM with a 500 RPM/second ramp.

(ii) Bake the resist layer over a hot plate at 115 °C for 70 s.

(iii) Expose the resist through a circular mask placed atop glass using a SUSS MA-6 mask aligner for 1 m.

(iv) Develop in CD-26 for 1 m.

Approximate step-time is 1 h for 8 samples.

***5a. Add a layer of oil within the developed circular photo-resist area.*** Early trials showed that small pinholes appear within the developed photoresist that result in short-circuiting the top and bottom contacts once the top AZO layer deposited. We avoid this by depositing oil inside the circular area of the developed resist to serve as a barrier that prevents etching of a-Si:H within the circle. Approximate step-time is 1 h for 8 samples.

***5b. Use $XeF_2$ to etch away the a-Si:H outside of the circular center.*** A flow of $XeF_2$ gas was used to etch a-Si:H where exposed outside the central circle, thereby exposing the cell bottom contact. This was done by flowing 10 pulses of the gas for 20 s each. Approximate step-time is 30 m for 3 samples.

***5c. Cleaning off the oil and photoresist.*** In this step, we remove the oil and photoresist used in Steps 5 – 7. We first squirt acetone on each sample individually and dip the samples in a fresh batch of acetone 2 times in a row, followed by dipping in methanol, and then IPA. Special care is taken to clean the tweezers used between each dipping step. Approximate step-time is 30 m for 4 samples.

## 6. Deposit top-contact AZO using ALD

We deposit the top AZO contact following the recipe used in Step 5 but at a temperature of 130 °C. Approximate step-time is 9 h.

## 7. Deposit Au using the sputtering tool

We sputter a 245-nm-thick layer of gold on the cells from Step 9 that serves as a top and bottom contact. The sputtering conditions are as follows:

Deposition time = 8 m; deposition height = 36; deposition power = 50% (150 W); deposition pressure = 4 mTorr; Ar flow = 40 SCCM

Approximate step-time is 30 m.

## 8. Photolithography and wet etching to define the Au contacts

In this step, we etch the excess gold and leave intact the top and bottom metal contacts as follows:

(i) Spin-coat S1818 Shipley photoresist for 5 s at 500 RPM with a 100 RPM/s ramp followed by a 45-s spin at 2000 RMP with a 500 RPM/s ramp.

(ii) Bake the resist layer over a hot plate at 115 °C for 70 s.

(iii) Expose the resist through a mask using a SUSS MA-6 mask aligner for 1 m.

(iv) Develop in CD-26 for 1 m.

(v) Wet etch the Au material in an electronics-grade Au etching solution.

Approximate step-time is 3 h for 8 samples.

## 9. Photolithography and wet etching to disconnect the top and bottom AZO contacts

In this step, we aim to etch an opening in the top AZO contact, thereby disconnecting the top and bottom contacts of the cell. The remaining top and bottom AZO contacts are in direct connection with the Au strips created in Step 8.

(i) Spin coat S1818 Shipley photoresist for 5 s at 500 RPM with a 100 RPM/sec ramp followed by a 45-s spin at 2000 with a 500 RPM/second ramp.

(ii) Bake the resist layer over a hot plate at 115 °C for 70 s.

(iii) Expose the resist through a mask using a SUSS MA-6 mask aligner for 1 m.

(iv) Develop in CD-26 for 1 m.

(v) Wet-etch the AZO material in an electronics-grade Al etching solution.

Approximate step-time is 3 h for 8 samples.

## 10. Deposition of the dielectric spacer and front mirror

In this step, we deposit a silica spacer (of thickness 2, 4, or 10 microns) and a second dielectric Bragg mirror on the solar-cell active area via e-beam evaporation at 200 °C. This dielectric Bragg mirror is partially reflective and comprises 6 alternating layers (3 bilayers) of $SiO_2$ and $TiO_2$ of thicknesses 129 nm and 86 nm, respectively. Approximate step-time is 13 h per run.

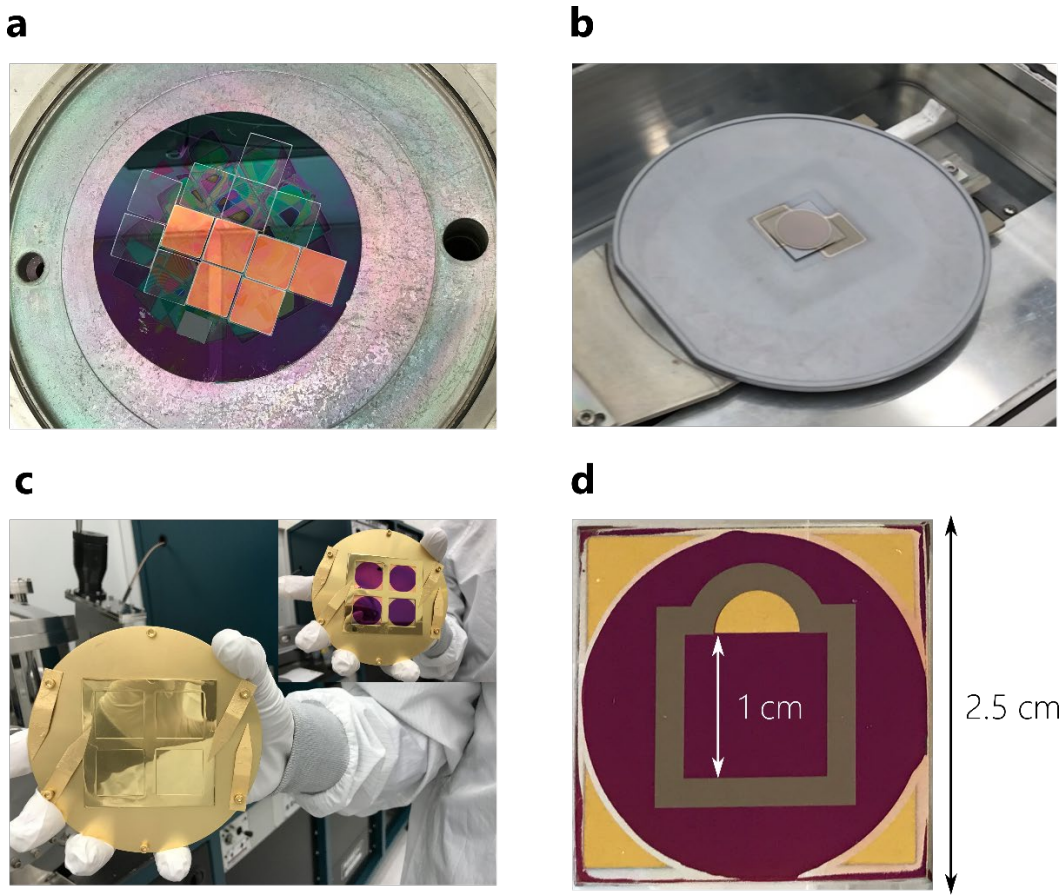

**Figure S4 | Solar cell fabrication at Harvard CNS Labs. a**, Samples in the ALD tool after AZO deposition (step 3). **b**, Sample in PECVD tool for a-Si:H deposition (step 4). **c**, Sample after sputtering and Au deposition (step 7). The inset shows the samples after depositing the PIN-diode (step 6) before Au deposition. **d**, Optical micrograph of the finished cell with Au contacts (step 9).

## S.3 Refractive indices of the solar cell layers

**a. Refractive indices of the dielectric mirrors**

The multilayer back-reflector and the partially reflective front-mirror are formed of alternating layers of $SiO_2$ and $TiO_2$. The refractive indices of these two materials in the spectral range of interest are shown in Fig. S5.

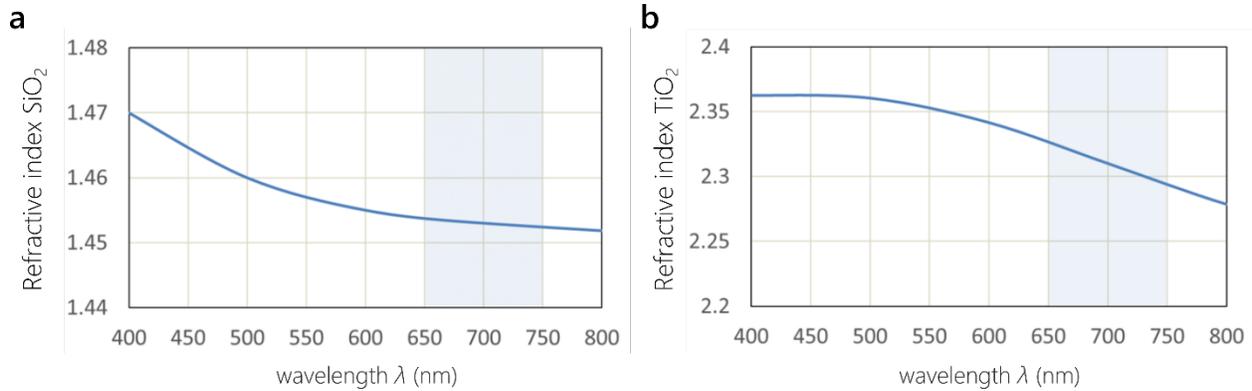

**Figure S5 | Refractive indices of SiO₂ and TiO₂. a,** Measured refractive index of $SiO_2$ obtained via spectroscopic ellipsometry. **b**, Measured refractive index of $TiO_2$ obtained similarly to (**a**).

**b. Refractive indices of the CPA Fabry-Pérot cavity layers**

The refractive indices of the five materials involved in constructing the PIN-diode incorporated into CPA Fabry-Pérot cavity were measured on the Woollam M2000 variable-angle mapping spectroscopic ellipsometer using witness samples (on glass substrates) for each layer produced during the various deposition steps. The materials are the n-type, i-type, and p-type a-Si:H layers, and the front and back AZO contact layers. We plot the real and imaginary parts of the wavelength-dependent refractive indices of these layers in Fig. S6. These measurements were then used in the transfer-matrix calculations of the absorption through the CPA-cavity as plotted in Fig. 5a and Fig. 5c,d in the main text.

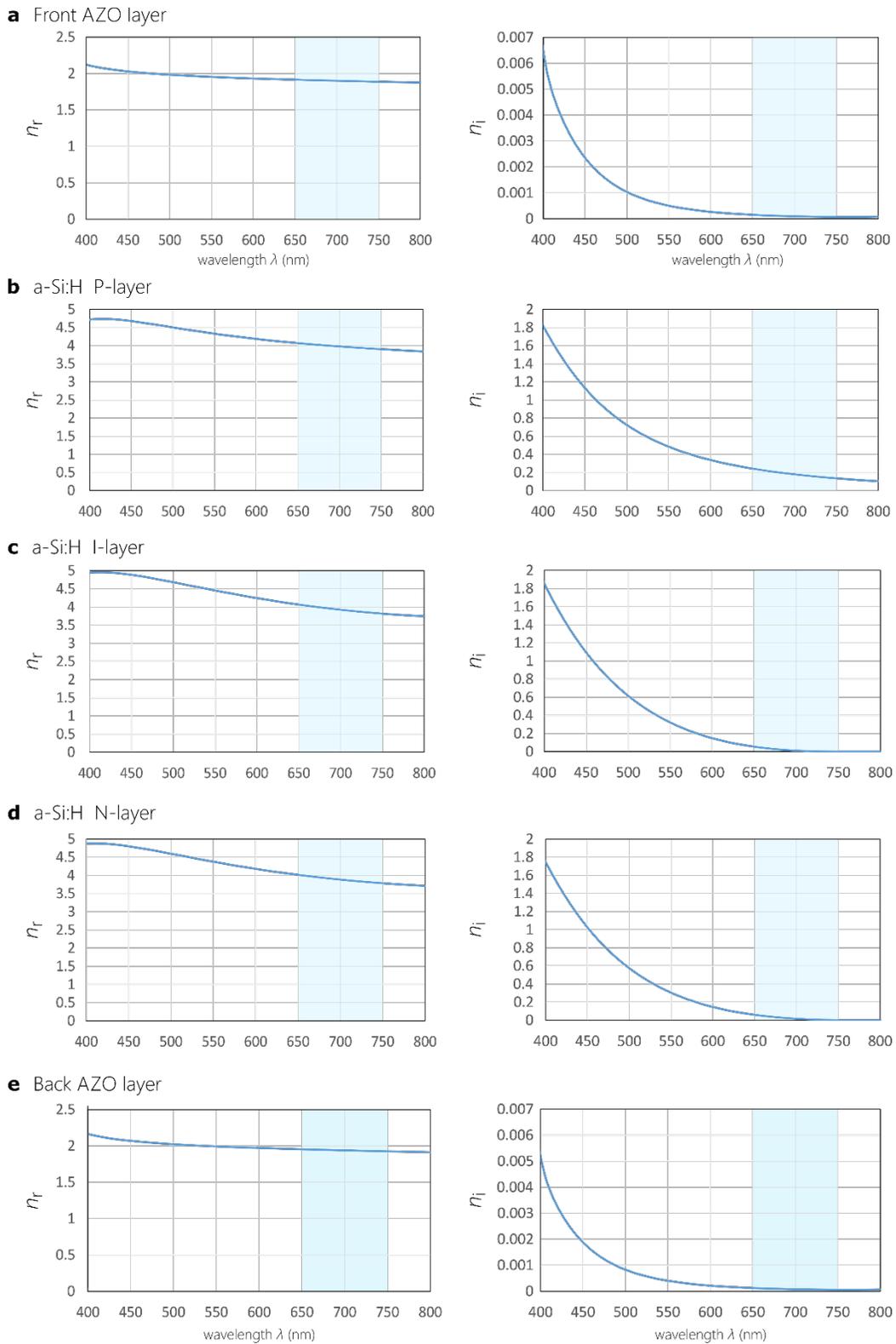

**Figure S6 | Refractive indices of the layers in the PIN-diode solar cell.** The wavelength-dependent real and imaginary parts of the refractive indices of the materials involved in constructing the PIN-diode solar cell from the top to bottom of the device: **a**, front AZO layer; **b**, P-layer; **c**, I-layer; **d**, N-layer; **e**, back AZO layer.

## S.4 Omni-resonance measurements configuration

Figure S7 depicts a schematic of the setup for omni-resonance that highlights the definition of the relevant angles for our analysis. The angularly dispersed light from the grating is directed to the cavity through the lens L$_1$. We take 710 nm as the central wavelength that defines the optical axis. The tilt angle of the sample $\psi$ is measured with respect to this optical axis. We define the angle $\gamma(\lambda)$, which is the diffraction angle with respect to the grating normal. The central wavelength $\lambda_c = 710$ nm is diffracted at $\gamma_o = \gamma(\lambda_c = 710$ nm$)$ and coincides with the optical axis. The angle any wavelength $\lambda$ makes with respect to this optical axis is $\gamma(\lambda) - \gamma_o$. This angle (and therefore the angular dispersion $\beta$) is boosted via the lens L$_1$ by a ratio $M = d_1/d_2$, where $d_1$ and $d_2$ are the distances from the grating to L$_1$ and from L$_1$ to the cavity, respectively; this is the inverse of the *spatial* magnification factor due to this single-lens imaging system. The incidence angle made by a wavelength $\lambda$ after the lens with respect to the optical axis is:

$$\varphi(\lambda) = \tan^{-1}\left\{\frac{d_1}{d_2}\tan(\gamma(\lambda) - \gamma_o)\right\}, \tag{S5}$$

where $\varphi_o = \varphi(\lambda_c = 710$ nm$) = 0$. The distances $d_1$ and $d_2$ are selected such that the illuminated spot on the grating is imaged onto the cavity. If the focal length of L$_1$ is $f$, then

$$d_2 = \frac{fd_1}{f-d_1}. \tag{S6}$$

When the cavity is oriented such that it is perpendicular to the optical axis, the angle of incidence of each wavelength is $\varphi(\lambda)$. Upon tilting the cavity by $\psi$, the angle of incidence with respect to the normal to the cavity is

$$\theta(\lambda) = \varphi(\lambda) + \psi. \tag{S7}$$

The angular dispersion $\beta$ produced for our dielectric spacer thicknesses and the required distances $d_1$ and $d_2$ in each case are:

2μm: $\beta = 0.410$ °/nm, $d_1 = 172.5$ mm, $d_2 = 29.2$ mm

4μm: $\beta = 0.355$ °/nm, $d_1 = 152.5$ mm, $d_2 = 29.9$ mm

10μm: $\beta = 0.310$ °/nm, $d_1 = 132.5$ mm, $d_2 = 30.8$ mm

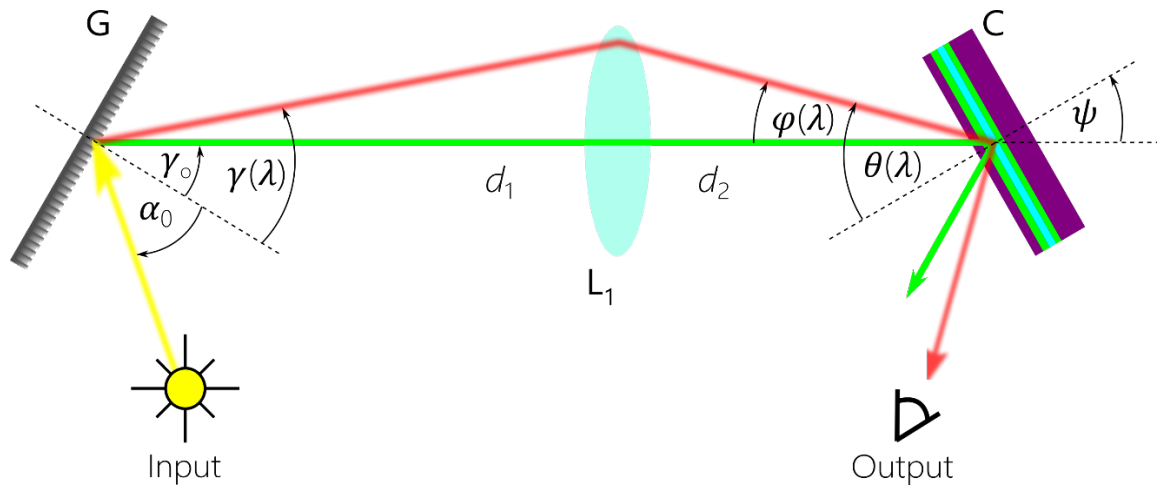

**Figure S7 | Schematic of the configuration of the grating and cavity to highlight the definitions of the various relevant angles.** $\alpha_0$ and $\gamma(\lambda)$ are measured with respect to the normal to the grating. The optical axis (shown in green) coincides with $\gamma_0 = \gamma(\lambda_c = 710 \text{ nm})$. $\varphi(\lambda)$ is measured with respect to the optical axis, while $\theta(\lambda)$ is measured from the normal to the cavity: $\theta(\lambda) = \varphi(\lambda) + \psi$, where $\psi$ is the tilt angle of the cavity with respect to the optical axis.

## S.5 Measurements of the angular dispersion

Figure S8a-c depicts the optical setups used in measuring the angular dispersion in the three configurations shown in Fig. 7a-c in the main text. Technical details of the setup are described in the Methods section in the main text. With respect to the last configuration in Fig. S8c (corresponding to Fig. 7c in the main text), the transmission diffraction gratings and prism used (Fig. S9a) lend themselves to the particularly simple arrangement shown in Fig. S9b. The thickness of the overall system will be further reduced by replacing the single prism shown in Fig. S9a with the microprism array shown in Fig. S9c. The size of the resulting configuration (Fig. S9d, corresponding to Fig. 7d in the main text) is vastly reduced.

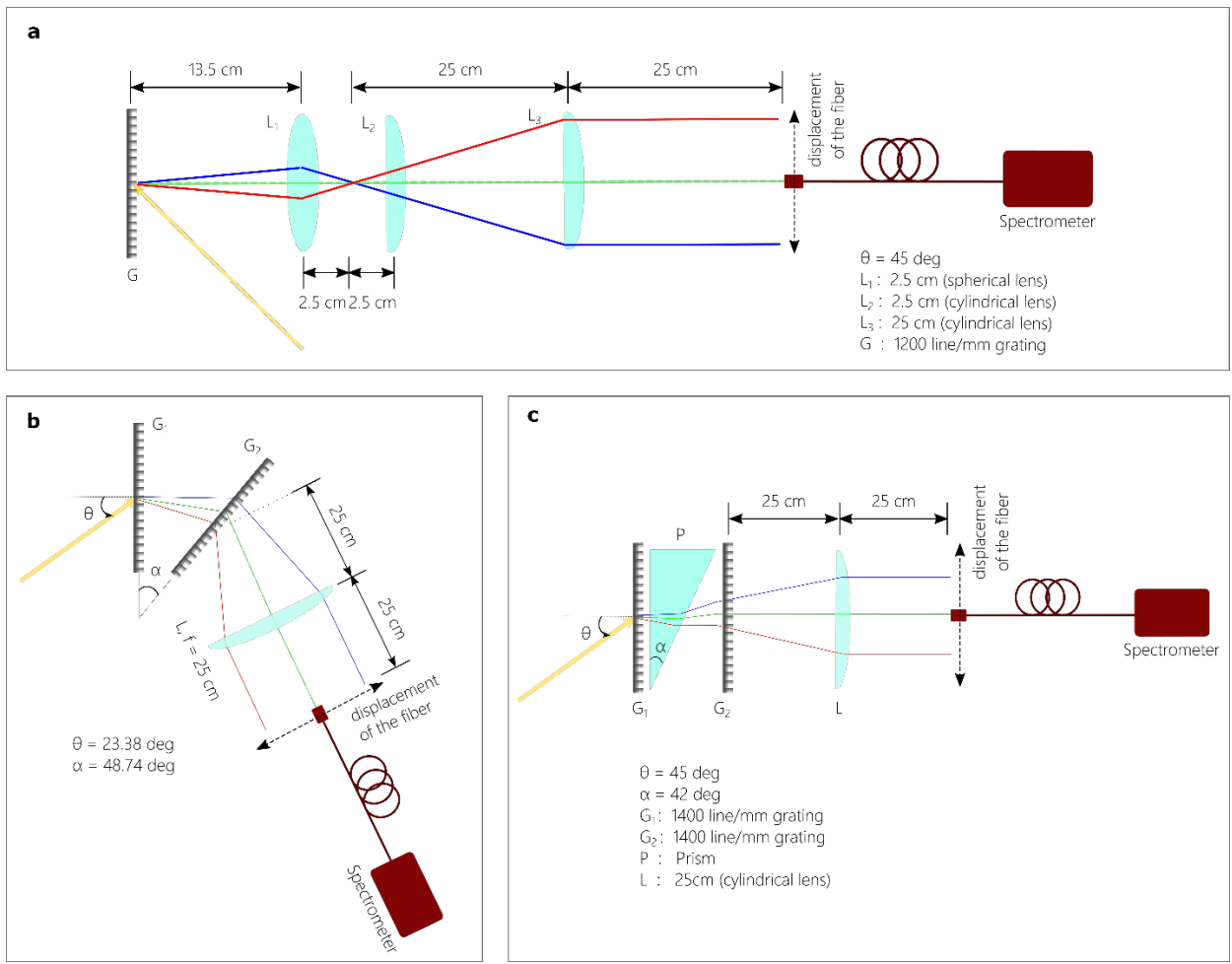

**Figure S8 | Optical setups for measuring angular dispersion. a**, Setup for measuring angular dispersion for the configuration shown in Fig. 7a in the main text. **b**, Setup for measuring angular dispersion for the configuration shown in Fig. 7b in the main text. **c**, Setup for measuring angular dispersion for the configuration shown in Fig. 7c and Fig. 7d in the main text.

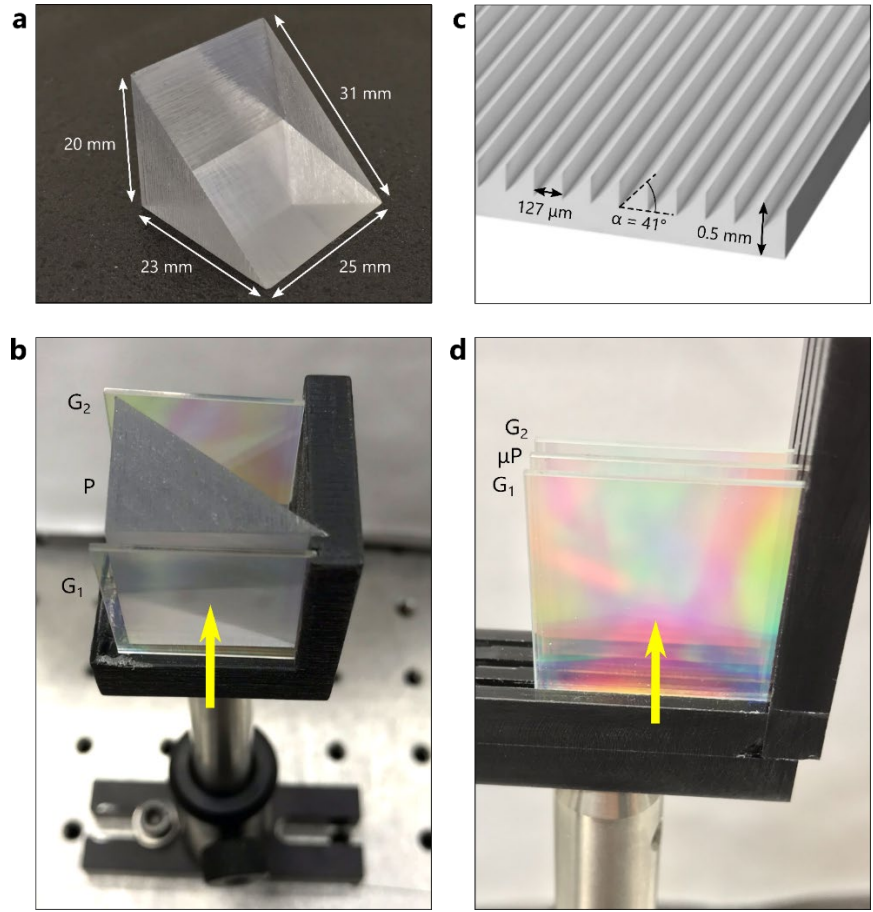

**Figure S9 | Single-prism and microprism-based arrangement for introducing angular dispersion in broadband light**. **a**, Photograph of a polymer prism fabricated by 3D-printing and polished to near-optical quality. **b**, Arrangement for introducing angular dispersion comprising a prism (P) sandwiched between two parallel transmissive diffraction gratings ($G_1$ and $G_2$) each of area 25×25 mm². **c**, Sketch of a polymer microprism array with significantly reduced thickness with respect to the single prism in (**a**) while introducing the same deflection angle. **d**, Arrangement for introducing the same angular dispersion achieved in (**c**) using a 0.5-mm thick microprism sheet (μP) arranged in a flat planar configuration between two parallel transmissive diffraction gratings ($G_1$ and $G_2$) as in (**b**). Note the reduced thickness of the whole arrangement. The yellow arrows in (**b**) and (**d**) indicate the incidence port.

# S.6 Solar cell characterization

The goal of this work is test the impact of broadband CPA on a solar cell, and not the demonstration of a record bare solar cell. We present here the electrical characterization of the bare solar cell (without the CPA cavity and light preconditioning system) under simulated solar irradiation of AM 1.5 G. It has an efficiency of ≈ 1% that is limited by the resistance of the transparent AZO contacts.

Bare Cell     Bottom AZO :   57 Ω/sq, Thickness = 312 nm, $n = 1.97$
                 Top AZO     : 180 Ω/sq, Thickness = 315 nm, $n = 2.12$

| $V_{oc}$ (V) | $I_{sc}$ (mA) | $J_{sc}$ (mA/cm²) | $I_{max}$ (A) | $V_{max}$ (V) | $P_{max}$ (mW) | Fill Factor | Efficiency |
|---|---|---|---|---|---|---|---|
| 0.595 | 5.03 | 5.03 | 2.824 | 0.33634 | 0.95 | 31.75 | 0.95 |

| $R$ at $V_{oc}$ | $R$ at $I_{sc}$ | Power (mW) | $R_{shunt}$ (Ω) | Cell Temp. start | Cell Temp. end | Exposure | Time | Date |
|---|---|---|---|---|---|---|---|---|
| 68.71 | 368.4 | 1.752 | NaN | NaN | NaN | 12.545 | 12:58:33 | 4/4/2018 |

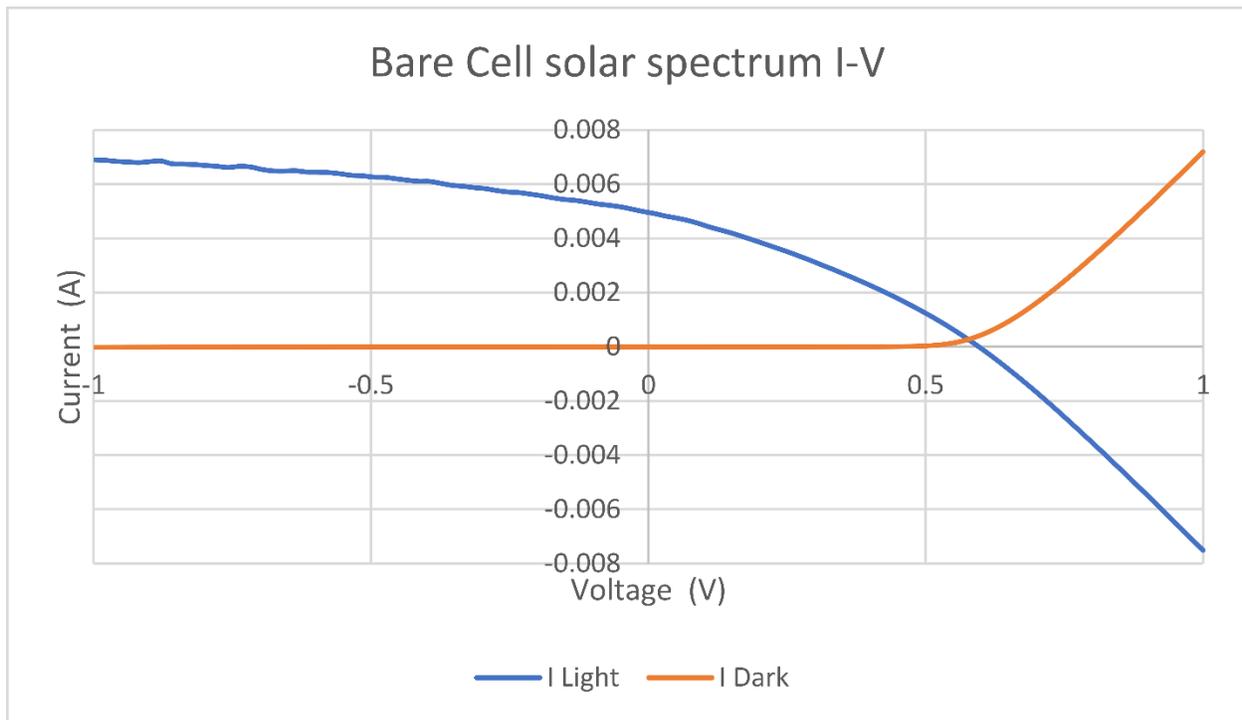

**Figure S10 | I-V characteristic of the bare solar cell under solar simulator irradiance.** Vertical axis is current in units of A (amperes). Horizontal axis is voltage in V (volts).